\documentclass[twocolumn]{aastex631}
\let\splitbox\relax
\usepackage{makecell}
\usepackage{csquotes}

\newcommand\pmra{$\mu^*_\alpha$}
\newcommand\pmdec{$\mu_\delta$}

\newcommand{\n}[1]{{\color{red} #1}}

\usepackage{verbatim}
\usepackage[T1]{fontenc}
\usepackage{braket}
\usepackage{adjustbox}
\usepackage{xpatch}

\usepackage{hyperref}
\usepackage{float}
\usepackage{graphicx}	
\usepackage{amsmath}	
\usepackage{booktabs,multirow}
\usepackage{soul}
\usepackage[table]{xcolor}
\usepackage{changepage,threeparttable}
\usepackage[shortlabels]{enumitem}
\newcommand{\data}{\ensuremath{M_\mathrm{data}}}
\newcommand{\modrv}{\ensuremath{M_\mathrm{3Dmodel}}}
\newcommand{\modpm}{\ensuremath{M_\mathrm{2Dmodel}}}

\begin{document}

\title{A Statistical Framework to Identify Kinematically Outlying LMC Globular Clusters and Implications for the LMC's Dark Matter Profile}

\correspondingauthor{Tamojeet Roychowdhury}
\email{tamojeet@berkeley.edu}

\author[0009-0003-9906-2745]{Tamojeet Roychowdhury}
\affiliation{Department of Astronomy, University of California Berkeley, Berkeley, CA 94704, USA}
\affiliation{Department of Electrical Engineering, Indian Institute of Technology Bombay, Mumbai, 400076, India}

\author[0009-0002-3202-1335]{Navdha}\affiliation{Perimeter Institute for Theoretical Physics, 31 Caroline Street North, Waterloo, ON, N2L 2Y5, Canada }
\affiliation{Department of Physics, Indian Institute of Technology Bombay, Mumbai, 400076, India}

\author[0009-0009-0158-585X]{Himansh Rathore}
\affiliation{Department of Astronomy and Steward Observatory, University of Arizona, 933 North Cherry Avenue, Tucson, AZ 85721, USA}

\author[0000-0002-7134-8296]{Knut A.G. Olsen}
\affiliation{NSF National Optical-Infrared Astronomy Research Laboratory, 950 North Cherry Avenue, Tucson, AZ 85719, USA}

\begin{abstract}
The LMC's Globular Clusters (GCs) bring a novel opportunity to understand the LMC’s assembly history and dark-matter (DM) properties, provided the kinematically outlying GCs can be reliably identified. However, traditional Energy – Angular Momentum space diagnostics fail because of large uncertainties on the GC velocities. In this work, we develop a new, robust statistical framework for identifying kinematically outlying LMC GCs, by using their Gaia-DR3 Proper Motions (PMs) combined with archival Line-of-Sight (LoS) velocity measurements. We use the difference between a GC's velocity vector and the average velocity vector of the field red clump stars as a metric for quantifying a GC's kinematic peculiarity. We account for the velocity measurement uncertainties and the LMC's intrinsic velocity dispersion. We find 5 GCs to be kinematically outlying based on PM differences alone, and additional 6 GCs if LoS velocity information is also used. Majority of the GCs with outlying PMs are clustered at a distance of $3-4$~kpc from the LMC center. The inclusion of outlying GCs introduces a bias of upto 30\% in the LMC's enclosed mass estimates derived using GCs as dynamical tracers; caution must be exercised in choosing the GC sample for precisely determining the LMC's DM content. We discuss the possibility that the kinematically outlying LMC GCs may be located in the LMC's elusive stellar halo and/or they could be accreted from external galaxies. Our work thus motivates future spectroscopic follow-up of these outlying GCs.   
\end{abstract}

\keywords{\href{http://astrothesaurus.org/uat/903}{LMC (903)}; \href{https://astrothesaurus.org/uat/656}{Globular Clusters (656)}; \href{https://astrothesaurus.org/uat/602}{Galaxy kinematics (602)}; \href{http://astrothesaurus.org/uat/1043}{Astronomical methods (1043)}; \href{http://astrothesaurus.org/uat/416}{Dwarf galaxies (416)}}

\section{Introduction}

The LMC is the most massive ($\sim 10^{11}$ M$_\odot$, e.g. \citealt{Watkins_2024}) satellite galaxy of the Milky Way (MW), and is the nearest ($\approx 50$ kpc away, \citealt{Pietrzy_ski_2013}) galaxy with a well-formed disk \citep[e.g.][]{van_der_Marel_2002, Choi_2022, Dhanush2024}. Precision astrometry from Gaia \citep{GaiaEDR3_MagellanicClouds2021} and Hubble Space Telescope (HST) \citep{Kallivayalil_2013} have enabled the use of stellar kinematics and cluster kinematics as a valuable tracer of assembly histories of satellite galaxies like the LMC. Further, these rich kinematic data make the LMC a new laboratory for dark matter (DM) physics \citep[e.g.][]{Besla2019, Foote2023, Cordova-Reynoso2024, DeBrae2025, Rathore_2025b}. Three dimensional velocity vectors of the LMC's Globular Clusters (GCs)\footnote{While the term \enquote{globular cluster} is often used to refer only to those massive, spherical star clusters that are old and metal-poor, in this paper we adopt the definition from the \href{https://astrothesaurus.org}{Unified Astronomy Thesaurus}, which allows the inclusion of clusters of any age and metallicity.} are finally available \citep[][hereafter B22]{bennet2022kinematic}, which allows studying the LMC's assembly history and DM properties with the GC population. A detailed identification of kinematically typical and atypical LMC GCs is necessary for effectively utilizing the GC kinematics as a tracer of the LMC's assembly history and DM content. In this work, we build a statistical framework for identifying kinematically outlying LMC GCs.

The LMC has an abundant GC population, with $\approx$50 GCs identified till date, including 15 old ($\geq 10$ Gyr in age) clusters \citep{Sarajedini_2024} and several young massive clusters \citep{Johsnon2001, Perina2010, Ahumada2019}. The LMC likely had a rich assembly history prior to its recent ($1-2$ Gyr, \citealt{Besla_2007}) infall into the MW halo \citep[e.g.][]{Donghia2008, Sales2011, Jethwa2016, Dooley2017, Jahn2022}. Further, in addition to the SMC, atleast 9 dwarf galaxies in the MW's neighborhood are either LMC satellites or have interacted significantly with the LMC over the last $\sim 1$ Gyr  \citep[e.g.][]{patel2020orbital, Erkal2020, Pace_2022, 10makarov2023}. Hence, a subset of the LMC GCs are possibly accreted from other galaxies. The kinematics of the GCs potentially encode whether they were born in-situ in the LMC, or brought-in from other galaxies.

Kinematics of the MW GC's have transformed our understanding of the MW's accretion history \citep[e.g.][]{Carraro1998, West2004, mackey2004, myeong2018, Helmi_2018, Massari_2019, Kruijssen_2019, Pagnini_2023} and DM profile \citep{Battaglia2005, Binney2017, Posti2019, Magnus-Correa2022}. At least $1/4$ of the MW GCs are accreted from other galaxies as opposed to forming in-situ \citep[e.g.][]{forbes2010}. The inference of the MW's merger history from its GCs relies on identifying kinematically outlying GCs that are likely brought in from other galaxies. Kinematically outlying MW GCs are usually identified from the Energy (E) - Specific Angular Momentum ($L_z$) distribution of the GC population \citep[e.g.][]{Helmi_2018, Belokurov_2024, 2024arXiv240806161C}. Motivated by this approach, in Figure \ref{fig:E-Lz}, we place the LMC GC population in the LMC's E - $L_z$ space. The large measurement errors on the LMC GC kinematics make it very challenging to ascertain kinematically outlying GCs with traditional diagnostics like the E - $L_z$ space. A new methodology is needed to identify kinematically outlying LMC GCs.  

Unlike the MW, most of the LMC GCs share the same rotation pattern as the LMC's disk and are consistent with being in the disk plane \citep{Schommer_1992, Mackey2003, McLaughlin2005, bennet2022kinematic}. Hence, comparing the GC velocity vectors to the velocity vectors of stars surrounding the GC in the LMC disk is a promising approach for identifying kinematically outlying GCs. However, this approach has several challenges that need to be addressed. 

The LMC's disk has a significant velocity dispersion (average $v_{rot}/\sigma = 2.9 \pm 0.9$, \citealt{van_der_Marel_2002}). The intrinsic velocity dispersion must be accounted for when comparing GC kinematics with the kinematics of the field stars. Further, spatially correlated systematic errors in the Gaia PM measurements \citep{GaiaEDR3_MagellanicClouds2021} and crowding induced incompleteness of star counts \citep{Rathore_2025} also hinders effective comparison between the GC kinematics and field star kinematics. The age differences between the GCs and the surrounding field stars also needs to be taken into account while interpreting the kinematic differences.

This work has three objectives:
\begin{itemize}
    \item Build a statistical framework to compare the kinematics of the LMC's GCs with the field LMC disk stars. In this framework, we shall take into account the LMC's intrinsic velocity dispersion and the statistical and systematic uncertainties of the Gaia measurements. The impact of diverse GC ages will also be assessed.
    \item Present a catalog of LMC GCs that are statistically significant kinematic outliers with respect to the field stars in the disk. Such GCs can have an external origin.
    \item Assess the reliability of using the LMC GC sample as a tracer of the LMC's DM content.
\end{itemize}

For building the statistical framework, we follow two different approaches to find GCs with significantly different kinematics from the field disk stars. The first approach (or the \enquote{data driven} approach) involves comparing the PM measurements of the GCs with the PMs of the field LMC stars. The second approach (or the \enquote{model driven} approach) involves using a LMC kinematic model to derive the expected velocities of the field stars at the location of each GC, and comparing the observed cluster velocities with the model derived velocities.

The data-driven approach allows us to directly take into account the intrinsic velocity dispersion of the LMC's disk while comparing GC kinematics to the field star kinematics. However, statistical and systematic uncertainties in the PM measurements can affect the inferences from the data-driven approach. The model-driven approach on the other hand allows us to evaluate a GC as a kinematic outlier independent of the limitations of the observational data. However, models of the LMC disk might not fully account for its disequilibrium \citep[e.g.][]{Choi_2022, Dhanush2024, Rathore_2025b}. Thus, using the data-driven approach and the model-driven approach simultaneously can help overcome the individual limitations of each of these approaches and yield a clean sample of outlying LMC GCs. 

Our manuscript is structured as follows. Section \ref{sec:data} presents the selection of LMC GCs and the sample of field disk stars around each GC. In section \ref{sec:framework}, we build a statistical framework for identifying kinematically outlyling GCs. Based on our statistical framework, we present a catalog of kinematically outlying GCs in section \ref{sec:results}. In section \ref{sec:discussion}, we place our results in context with previous works, investigate the reliability of the LMC GC sample for measuring the LMC's DM content, and assess the limitations and caveats of our analysis as well as future prospects. We conclude in section \ref{sec:conclusion}.

\section{Data Selection} \label{sec:data}

\subsection{Selection of LMC Globular Clusters}

We utilize the catalog provided by \citet[][hereafter B22]{bennet2022kinematic} for the properties of the LMC GCs. B22 provide 6-D phase space information - Right Ascension (RA), Declination (DEC), parallax, proper motion (PM) and radial velocity (RV) along with the associated errors for a sample of 30 LMC GCs. The PMs of the GCs have been obtained by combining astrometry from \textit{Gaia} DR3 and the Hubble Space Telescope (HST).

B22 carefully evaluated the membership probability of the candidate stars belonging to a GC using a maximum likelihood estimation algorithm. They define the likelihood with a Gaussian Mixture Model, which is constructed by combining the positions, PMs, parallaxes and the location of the stars in the color-magnitude diagram (CMD). Then, they performed iterative Gaussian profile fitting in parallax and PM space and rejected the outlying stars in each iteration, thereby obtaining a clean sample of stars belonging to a particular GC.

B22 obtained the PMs (and their errors) of the GCs as a whole by applying the \texttt{GetGaia} pipeline \citep{delpino2021, martinez_garcia2021} on the cleaned sample of member stars. For GCs with archival HST measurements, they used the \texttt{GaiaHub} pipeline \citep{del_Pino_2022} to perform photometric source detection and cross-matching to \textit{Gaia} sources. The final values of the GC properties used in this work are adopted from Table 2 of B22.

\subsection{LMC GC's in the Energy - Angular Momentum Space}
\label{sec:E-Lz}

The specific energy (E) -- specific angular momentum ($L_z$) parameter space has been widely utilized to identify kinematically outlying stellar populations in the Milky Way \citep[e.g.][]{Helmi_2018, Belokurov_2024,2024arXiv240806161C}. We apply the same technique on our set of GCs to infer if traditional diagnostics like the $E-L_z$ space provide any insights into significantly outlying LMC GC populations.

For the rest of this paper, we use $(r, \phi, z)$ to refer to a cylindrical coordinate system centered on the LMC and aligned with its disk plane. The LMC photometric center (RA, DEC = $81.28^\circ, -69.78^\circ$), disk inclination ($37.4^\circ$) and the coordinate transformation equations for converting the equitorial coordinates to $(r, \phi, z)$ are taken from \cite{van_der_Marel_2001} and \cite{van_der_Marel_2002}. A similar reference frame was adopted in the B22 catalog.

We take the values of $r$, $v_\phi$, and the uncertainty in $v_\phi$ [$\epsilon(v_\phi)$] for the GCs from B22, and compute the specific angular momentum: 
\begin{equation}
L_z = r v_\phi   
\end{equation}
\noindent and its error:
\begin{equation}
\epsilon(L_z) = r \times \epsilon(v_\phi)    
\end{equation}
The specific kinetic energy of the GCs is computed using the squared sum of residual velocities. The residual velocities are obtained after subtracting the LMC's systemic motion from the total 3-D velocity vectors of the GCs. For the LMC, we adopt (\pmra, \pmdec) = (1.910, 0.229) mas yr$^{-1}$ \citep{Kallivayalil_2013} and RV = 262.2 km s$^{-1}$ \citep{McConachie_2012}. These values for the LMC's systemic motion were also used by B22. The error in kinetic energy is computed as the sum $\sum_i v_i\epsilon_{vi}$, where $v_i$ is the residual velocity component (\pmra, \pmdec\ , RV) and $\epsilon_{vi}$ is the corresponding uncertainty in that component.

For computing the specific potential energy of the GCs, we model the LMC's dark matter (DM) halo potential as a Hernquist profile \citep{Hernquist_1990}, with a total halo mass of $1.8\times10^{11}M_\odot$ and a Hernquist scale radius of $23.1$ kpc \citep{patel2020orbital}. The LMC's disk potential is modeled as a Miyamoto-Nagai function \citep{Miyamoto_Nagai_1975}, with a total disk mass of $3\times 10^9M_\odot$, scale radius of $1.7$ kpc and scale height of $0.27$ kpc \citep{patel2020orbital}.  Using the $r$ and $z$ values of the GCs from the B22 catalog, we compute the specific potential energy of each GC. With a fixed coordinate system, the uncertainties in the radial distance from LMC centre ($r$) are small and the total energy uncertainties are dominated by the kinetic energy uncertainties for all GCs.

\begin{figure}
  \centering
\includegraphics[width=\columnwidth]{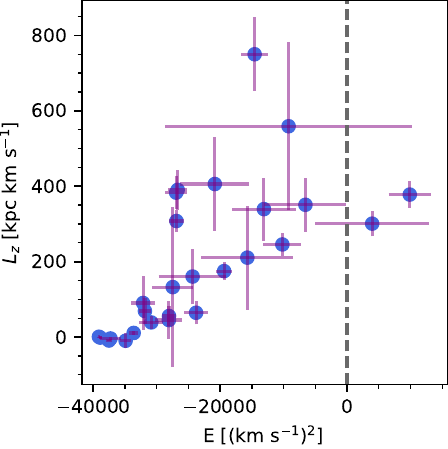}
\caption{The sample of LMC GCs used in our work is plotted in the specific energy (E, x-axis) v/s specific angular momentum ($L_z$, y-axis) space, with their $1\sigma$ errorbars. The coordinate frame is described at the start of section \ref{sec:E-Lz}. Given the large errorbars on most of the GCs, it is challenging to identify kinematic outliers with traditional diagnostics like the E-$L_z$ space, and a new framework is needed. The grey dashed vertical line marks $E = 0$, and GCs that reside to the right of it (NGC 2210, NGC 2159) are likely unbound to the LMC, making them obvious kinematic outliers. Of these, NGC 2210 is inconsistent with $E<0$ even within a $1-\sigma$ error.}
\label{fig:E-Lz}
\end{figure}

$E$ is given by the sum of the specific kinetic energy and the specific potential energy. We plot $E$ v/s $L_z$ along with $1\sigma$ errorbars in Figure~\ref{fig:E-Lz}. Given the large errorbars on $E$ and $L_z$, it is challenging to identify GCs that are kinematic outliers from this plot. Clustering of GCs in the $E - L_z$ space is also hard to identify given the large errorbars.  We note that two GCs, NGC 2210 and NGC 2159 have $E>0$ and hence are likely unbound to the LMC, and are obvious kinematic outliers. These two GCs have been considered outliers in previous works as well, like B22 and \cite{Watkins_2024} (we shall see later that our new statistical framework also flags both of these GCs as kinematic outliers). 

Note that in Figure~\ref{fig:E-Lz} we ignored uncertainties in $r$ and $z$, so the errorbars may be underestimated by upto a factor of 2. This makes this diagnostic even harder to use to separate kinematic outliers. The inability of the $E-L_z$ space to identify outlying GCs necessitates a framework that directly evaluates the differences between the GC kinematics and the kinematics of the field LMC disk stars. Further, to reliably assess any difference, we need statistically well-defined metrics that account for non-physical (like measurement uncertainties) and physical (non-zero velocity dispersion) agents that can lead to the GC kinematics being different from the field stars. Hence, in the subsequent subsections we introduce methods to compare the GC velocities to those of surrounding field stars directly.

\subsection{Selection of the Field Stars} \label{sec:fielddata}

The field stars associated with a particular LMC GC are chosen to be the red clump (RC) stars in the vicinity of that GC from the \textit{Gaia} DR3 catalog \citep{2016, Lindegren_2021, gaiacollaboration2022gaia}. The choice of RC stars as tracers of the field population is motivated by several factors: 
\begin{itemize}
    \item They are an old stellar population (age $>$ 1 Gyr, \citealt{Girardi2016}), and hence are good dynamical tracers of a galaxy disk. Several studies on the structure and kinematics of the LMC disk have used the RC population \citep[e.g.][]{Subramaniam2009, Choi2018, Choi_2022, Rathore_2025}. 
    \item They are similar in age to the dominant population of stars that constitute the GCs (ages between 2 - 12 Gyr, \citealt{He_2022}). However, while the RC populations consist of a range of ages, any one GC typically consists of stars having a narrow distribution of ages. We will assess the effect of age on the GC kinematics later in the manuscript. Also, it is likely that the relatively older populations would have the signatures of a disk that is kinematically altered over time through both secular evolution and interactions with other galaxies like the SMC. Our aim is to find GCs that are anomalous even with respect to this already altered disk population.
    
    \item They occupy a well-defined region in the CMD \citep{Castellani2000}, which makes them easy to select with \textit{Gaia} photometry.
    \item Kinematic models of the LMC based on red clump stars have already been developed \citet[][hereafter C22]{Choi_2022}. Hence, utilizing RC stars would ensure consistency when we compare the PMs of the GCs to the model derived velocities at the GC location.
\end{itemize}

We select RC stars in a surrounding annulus around each GC, using the following algorithm:

\begin{itemize}
    \item The central sky coordinates (RA \& Dec) of each GC are queried from SIMBAD \footnote{\url{https://simbad.cds.unistra.fr/simbad/sim-fid}}. If the GC center coordinates are $\alpha_0$ and $\delta_0$, and a target field star coordinate (from Gaia) are $\alpha$ and $\delta$, then we define:
    \begin{equation}
        R^2 = (\alpha-\alpha_0)^2\cos{\delta_0}^2+(\delta-\delta_0)^2
        \label{eq:ra-dec-radius}
    \end{equation}
    \noindent Here, $R$ represents the projected on-sky separation between the GC and a target field star.
    
    \item We retrieve all stars residing within an annulus defined by an inner radius $R_{in}$ and an outer radius $R_{out}$ with respect to the GC center (exact choices for $R_{in}$ and $R_{out}$ are explained later in section \ref{subsec:selection-radii}) satisfying: 
    \begin{enumerate}
        \item $1.0 \leq G_{BP} - G_{RP} \leq 1.3$
        \item $18.6 \leq G \leq 19.3$
        \item $(\mu_\alpha^* - 1.8593 )^2 + (\mu_\delta - 0.3747)^2 \leq 1.5^2$
    \end{enumerate}
     These selection criteria are motivated from \cite{Choi_2022} and \cite{Rathore_2025}. Here $G_{BP}$, $G_{RP}$ and $G$ are the magnitudes in the Gaia blue photometer, red photometer and the G-band. $\mu_\alpha^* = \mu_\alpha \cos{\delta}$ and $\mu_\delta$ are the proper motions in RA (declination corrected) and DEC respectively. The criteria (1.) and (2.) select the red clump region in the Gaia CMD. The criteria (3.) ensures that the PMs of the chosen stars are not significantly different from the systemic PM vector of the LMC. The resulting sample of stars were queried from \textit{Gaia} DR3 using the \texttt{astroquery.Gaia} Python package \footnote{\url{https://astroquery.readthedocs.io/en/latest/gaia/gaia.html}}.
    
    \item Finally, to limit contamination from the foreground MW stars, we use LMC membership probabilities as calculated by \citet[][hereafter JA23]{jimenez-arranz2023}. JA23 developed a supervised neural network classifier for determining LMC membership probabilities of Gaia sources. We find $\sim 82\%$ of the stars in this catalog to have a probability of being an LMC star in JA23 ($P_\mathrm{LMC}$) $<0.1$ or $>0.9$, which indicates that the stars in the MW foreground (with low $P_\mathrm{LMC}$) and stars in the LMC (with high $P_\mathrm{LMC}$) are clearly separated. Following B22, we utilized the optimal selection criteria of JA23 to prioritize completeness of LMC stars, keeping only those Gaia sources having $P_\mathrm{LMC}>0.52$. This selection retains $47.7\%$ of the total stars in the region.
\end{itemize}

\subsubsection{Inner \& Outer Radii for field Star Selection}
\label{subsec:selection-radii}

In this section, we explain in detail our choice for the inner and outer radii of the annuli we defined in section \ref{sec:fielddata} for selecting stars surrounding a particular GC. The value of $R_{in}$ is taken to be 0.05 degree for all GCs. Adopting an LMC distance of 50 kpc from \cite{Pietrzy_ski_2013}, this angular extent corresponds to a spatial extent of 45 pc. The half-light radii of most globular clusters are about 10 pc or less \citep{van_den_Bergh_2008} while $r_{90}$ (radius enclosing 90\% luminosity) values are  $\sim20$ pc \citep{Werchan_2011}. Tidal radii (radius beyond which the galactic gravitational field is stronger than the GC's own field) are 20-50 pc for some representative LMC GCs studied in \cite{Piatti_2018}. Hence, our choice of $R_{in}$ ensures an almost exclusive selection of stars outside of the GC. 

We decide $R_{out}$ based on the number of stars present in the annulus. We initially set $R_{out}$ at 0.15$^\circ$, which yields between 2000 and 3500 field stars for most of the GCs. To ensure uniformity on the number of field stars for each GC, we increase or decrease the value of $R_{out}$ in steps of $0.01^\circ$ so as to get a count between 2000 and 3500. This range of star counts is necessary for reliable estimation of the mean PM and the PM dispersion of the field star sample, as justified in the paragraphs below.

We use the algorithm developed by \cite{Vasiliev_2018, Vasiliev_2019a, Vasiliev_2019b} (hereafter referred to as the V19 code) to estimate the mean PM (with uncertainty) and the PM dispersion of the field star sample. The aforementioned cited works have used this code for analyzing the PMs of the GCs in the Milky Way and the LMC. The V19 code ensures optimal performance with astrometric data at the LMC distance, particularly for crowded fields. We refer the reader to Appendix A of \cite{Vasiliev_2019a} for the details of the code, like the treatment of systematic errors in the observed astrometry and the fitting procedure via likelihood maximization to obtain the PM mean and the PM dispersion of a sample of Gaia stars. 

Briefly, the $2N$ proper motion components of a set of $N$ stars are assumed to be drawn from a $2N$-dimensional Gaussian. The $N$ mean values for the \pmra\ and the \pmdec\ components are assumed to be the same, and denoted as $m_\alpha^*$ and $m_\delta$ respectively. The diagonal elements of the covariance matrix used to define the Gaussian have values equal to the quadrature sum of the intrinsic PM dispersion and the measurement uncertainty for the PM component of a star. The off-diagonal element for \pmra\ and \pmdec\ of the same star is equal to the correlated error in the measurement of \pmra\ and \pmdec. The off-diagonal element corresponding to the PM components of two different stars is taken to be 0. This covariance matrix is then used in a log-likelihood maximization procedure to find the most probable values for the mean PM and the intrinsic PM dispersion. For calculating the PM uncertainties, the Hessian matrix with respect to the model parameters (mean and dispersion of GC PM) is computed.

The V19 code works best when the count of stars in the region of interest is $\sim 1000$. The method results in progressively higher errors on the returned values (mean PM and dispersion) as the number of stars decreases. For this reason, any GC whose surrounding region has $<200$ stars has too high errors ($\geq 0.05$ mas yr$^{-1}$) on the returned PM values, and is consequently rejected from further analysis. The upper error threshold that we adopt ($\geq 0.05$ mas yr$^{-1}$) is defined by observing the typical values of PM differences between the GC and field stars, which are $\sim 0.1$ mas yr$^{-1}$. Hence, if the measurement error in the mean PM is greater than 0.05 mas yr$^{-1}$, we cannot interpret the kinematic differences between the GC and field PMs. Low counts of field stars primarily happen for the outermost GCs residing in sparser regions of the LMC. 

On the other hand, if the star counts in the annuli for the field stars becomes larger than 5000, then the covariance matrix for the two PM components will be of dimensions $10^4\times 10^4$. Inverting this matrix for the likelihood computation (of the PM distribution) is computationally infeasible. The V19 code performs optimally when the star counts are between 2000 and 3500 (E. Vasiliev {\em priv. comm.}). Further, we will show later that the data-derived proper motion dispersions are systematically higher than the model-derived dispersions at all GC locations, indicating that our choice of sample size does not cause an underestimate of the dispersion.

In addition to the above criteria of star counts, we impose a lower limit and an upper limit on what $R_{out}$ can be at 0.1$^\circ$ and 0.2$^\circ$ respectively. The lower limit is chosen so that $R_{out}$ is at least twice $R_{in}$. This annulus area ensures that the region is small enough to capture the local mean PM in the field around the cluster, while enclosing enough stars to minimize measurement uncertainties. The upper limit corresponds to roughly $180$ pc at the LMC distance, which is between $3$ and $5$ times the typical tidal radius of LMC GCs \citep{Piatti_2018}. Further, beyond an on-sky extent of $0.2^\circ$, the spatially-correlated systematic errors between Gaia PMs can no longer be assumed to be constant (P. McMillan {\em priv. comm.}, see also \citealt{Gaia_2018}).

Our field sample of RC stars, therefore is not completely devoid of possible members of the GC itself. However, any contaminants in the sample would only lower the PM difference between the GC and the field stars. GCs having a high difference with our field sample will likely have an even higher difference if a more cleaner sample is selected. Our conclusions will still hold.

\section{Statistical Framework for Identifying Outlying LMC Globular Clusters} \label{sec:framework}

We compute the difference between the GC velocities and the velocities of the field stars. To establish statistical significance, we compare the difference with measurement errors as well as the intrinsic velocity dispersion of the field stars. The velocities for the field stars are calculated using two methods. The first method involves directly calculating the velocities using the Gaia DR3 measurements. The second method involves estimating the velocities using a kinematic model of red clump stars developed by C22.

The GC velocities and their errors are directly adopted from the B22 catalogue. The catalogue reports the mean PM in RA and DEC and their corresponding uncertainties for each GC.

\subsection{Field Star Velocities Using \textit{Gaia} DR3 Dataset}
\label{subsec:vasiliev}

We use the final sample of field stars as selected in section \ref{sec:fielddata}. We compute the the resultant mean PM along RA and DEC, their corresponding uncertainties, and the intrinsic PM dispersion using the V19 code.

\subsection{Field Star Velocities Using an LMC Kinematic Model}
\label{subsec:choi_model}

We use the kinematic model for the LMC's RC disk developed by C22 based on the formalism of \cite{van_der_Marel_2002}. We note an important discrepancy in the coordinate system assignment -- the adopted LMC center are different in the B22 catalog and the C22 model. B22 uses the LMC photometric center from \cite{van_der_Marel_2002}, whereas C22 determine the LMC kinematic center using PMs of RC stars. Further, the inclination of the disk is also different (taken $37^\circ$ in B22, inferred to be $23^\circ$ from model fitting in C22). These differences arise because the LMC's center depends on the stellar population used, and whether PM information is used in computing the center or not \citep{Rathore_2025}. To ensure the discrepancy in the LMC's center and disk inclination does not affect our statistical framework, we calculate the field RC velocities in the C22 adopted coordinate frame, and then transform all velocities back to PMs and RVs before performing any statistical computations. Since the definition of the PMs and RVs is LMC frame independent, our conclusions are robust to the choice of reference frame.

We now briefly describe the C22 kinematic model in their adopted coordinate system (here referred as $\tilde r, \tilde\phi, \tilde z$ to distinguish it from the B22 coordinate system $r, \phi, z$). Velocities in the $r'$ and $z'$ direction are taken to be 0. The tangential velocity profile as a function of distance from the LMC center is given by 
\begin{align}
    v(\tilde r) = \begin{cases}
        v_0\left( \frac{\tilde r}{\tilde r_0} \right),& \tilde r<\tilde r_0 \\
        v_0, & \tilde r>\tilde r_0
    \end{cases}
\end{align}
This equation results in a velocity profile that increases linearly until the scale radius $\tilde r_0$, after which it flattens to the constant value $v_0$. All of $v_0, \tilde r_0$ and $\eta = \tilde r_0 / d_{\mathrm{LMC}}$ are inferred from model fitting to RC PMs. 

We compute the model-based mean PMs and RVs for the field stars at the GC locations by supplying the best-fit parameters in C22 to the velocity equations (eq. 1--36) in \cite{van_der_Marel_2002}. These are then converted back to the observed frame of PM and RV before further computations. The computation of the PM dispersion at each GC location is detailed in section \ref{subsec:model-error-sigma}. 

\subsection{Constructing Statistical Metrics for Velocity Differences}
\label{sec:statmetrics}

The net PM difference between the GC and its field stars is calculated as a 2D vector, with components $\Delta \mu_\alpha^*$ and $\Delta \mu_\delta$ for difference in RA (pre-corrected with $\cos \delta$) and DEC respectively. The net difference becomes: 
\begin{equation}
\Delta_{\rm{pm}} = \sqrt{\Delta \mu^{*2}_\alpha + \Delta \mu^2_\delta}    
\end{equation}
Let the error in $\Delta_{\rm{pm}}$ be $\epsilon_{\rm{pm}}$ and the intrinsic velocity dispersion of the field stars be $\sigma_{\rm{pm}}$. The calculation for $\epsilon_{\rm{pm}}$ and $\sigma_{\rm{pm}}$ are different for data-obtained field velocities (section \ref{subsec:vasiliev}) and model-obtained field velocities (section \ref{subsec:choi_model}), and are elaborated in the following subsections (section \ref{subsec:data-error-sigma} and section \ref{subsec:model-error-sigma}).

We construct two metrics to check for the statistical significance of the PM difference ($\Delta_{\rm{pm}}$). We call these metrics $Q_{\rm{err}}$ and $Q_{\rm{disp}}$, which are defined as:
\begin{equation}
     Q_{\rm{err}} = \frac{\Delta_{\rm{pm}}}{\epsilon_{\rm{pm}}} \label{eq:q1-def}  
\end{equation}
\begin{equation}
     Q_{\rm{disp}} = \frac{\Delta_{\rm{pm}}}{\sigma_{\rm{pm}}} \label{eq:q2-def}
\end{equation}
$Q_{\rm{err}}$ quantifies how significant is the PM difference between a GC and its field stars when accounting for the measurement error. $Q_{\rm{disp}}$ quantifies how significant the difference is when accounting for the intrinsic velocity dispersion at that location. 

We outline three different scenarios for calculating $Q_{\rm{err}}$ and $Q_{\rm{disp}}$ for each GC, to ensure the best selection of GCs with a statistically significant velocity difference from its surroundings.

\subsubsection{Using Gaia DR3 Measurements of the Field Stars}
\label{subsec:data-error-sigma}

Let $\epsilon_{\alpha}(\rm{GC})$ and $\epsilon_{\alpha}(\rm{field} )$ denote the PM errors in RA for the GC and the field stars respectively. Let $\epsilon_{\delta}(\rm{GC})$ and $\epsilon_{\delta}(\rm{field} )$ denote the PM errors in DEC for the GC and the field stars respectively. The errors for the GC [$\epsilon_{\alpha}(\rm{GC})$ and $\epsilon_{\delta}(\rm{GC})$] are obtained from the B22 catalogue, and the errors for the field stars [$\epsilon_{\alpha}(\rm{field} )$ and $\epsilon_{\delta}(\rm{field} )$] are obtained with the computation in section \ref{subsec:vasiliev}. The final error in the net PM difference $(\epsilon_{\rm{pm}})$ is calculated by a Gaussian error propagation as follows:
\begin{align}
    \epsilon_\alpha &= \sqrt{\epsilon_{\alpha}(\rm{GC})^2 + \epsilon_{\alpha}(\rm{field} )^2} \\
    \epsilon_\delta &= \sqrt{\epsilon_{\delta}(\rm{GC})^2 + \epsilon_{\delta}(\rm{field} )^2} \\
    \epsilon_{\rm{pm}}^2 &= \left( \frac{\Delta\mu_\alpha^*}{\Delta_{\rm{pm}}} \right)^2 \epsilon_\alpha^2 + \left( \frac{\Delta\mu_\delta}{\Delta_{\rm{pm}}} \right)^2 \epsilon_\delta^2
\end{align}

Since the radial density of stars drops sharply in GCs and the $R_{in}$ used for the surrounding annulus is of the typical tidal radius value, we expect contamination from the GC stars in the field sample to be minimum. This ensures the errors on the PMs of the GC population and the field star population can be assumed to be nearly independent, justifying Gaussian propagation. The dispersion values $(\sigma_{\rm{pm}})$ obtained in section \ref{subsec:vasiliev} are directly used to compute $Q_{\rm{disp}}$ for each GC. 

The aforementioned scenario of using the \textit{Gaia} DR3 measurements of field stars to compute the GC--field PM differences is henceforth referred to as the \enquote{\data\ scenario}. 

We note that in the above treatment, systematic errors in the Gaia PM measurements have not been considered. The systematic errors are spatially correlated (as shown in \citealt{Lindegren_2021}). We have further verified the presence of this spatial correlation by examining the PM systematic error maps of the LMC field on an $8^\circ \times 8^\circ$ region (received from Paul McMillan, {\em priv. comm.}, see also Figure 17 of \citealt{Gaia_2018}). We do not expect the spatially correlated systematic errors to affect our statistical framework, as justified in the paragraph below.

The values reported for the GC PMs in the B22 catalogue are not corrected for the aforementioned systematic errors. Mathematically, if $\mu_\delta(\rm{GC})$ is the truth value of the GC PM along the DEC direction, the value reported in B22 is actually $\mu_\delta(\rm{GC})+\varepsilon_{\rm{sys}}$, where $\varepsilon_{\rm{sys}}$ is the systematic error  for $\mu_\delta$ at the GC's spatial location. Since the systematic error is spatially correlated at length scales $<0.2^\circ$ (which is the maximum outer radius we use for the field star selection), our method to find the PMs of field stars (section \ref{subsec:vasiliev}) returns the value $\mu_\delta(\rm{field} )+\varepsilon_{\rm{sys}}$, where $\mu_\delta(\rm{field} )$ is the truth value of the mean PM of the field stars. When we compute the difference ($\Delta\mu_\delta$) between the GC PM and the field star PM, the systematic term ($\varepsilon_{\rm{sys}}$) cancels out. In this paragraph, we have used the DEC component just for illustration, but a similar reasoning applies to the RA component.

\subsubsection{Using a 2D Kinematic Model for Field Star Velocities}
\label{subsec:model-error-sigma}

The velocity field of the field stars at the GC coordinates are derived as given in section \ref{subsec:choi_model}. In the model based approach, the error on the velocities of the field stars is taken to be 0. Hence, the only contribution to $\epsilon_{\rm{pm}}$ is from the GC's PM measurement errors, which are obtained by Gaussian error propagation from the RA and DEC components:
\begin{align}
    \epsilon_{\rm{pm}}^2 &= \left( \frac{\Delta\mu_\alpha^*}{\Delta_{\rm{pm}}} \right)^2 \epsilon_\alpha(\rm{GC})^2 + \left( \frac{\Delta\mu_\delta}{\Delta_{\rm{pm}}} \right)^2 \epsilon_\delta(\rm{GC})^2
\end{align}
where the terms denote the same quantities as section \ref{subsec:data-error-sigma}. The above equation is used to calculate $Q_{\rm{err}}$.
A local velocity dispersion model (i.e. velocity dispersion as a function of $r, \phi, z$) has not been developed for the LMC RC population in prior literature, and is currently beyond the scope of this work. However, to get an estimate of the velocity dispersion that we can use for our model driven approaches (outlined below), we utilize the velocity dispersion curve for the LMC red giant stars formulated in \cite{Vasiliev_2018}. The coordinate reference frame in \cite{Vasiliev_2018} (adopted LMC center and disk inclination) is roughly the same as in B22, and the minor differences do not affect the overall conclusions of our analysis. In \cite{Vasiliev_2018}, velocity dispersions are reported along the $(r,\phi)$ directions ($\sigma_{R,\phi}$) and along the $z$ direction ($\sigma_{z}$) separately, as a function of distance from the LMC center ($r$). 

To calculate the PM dispersion in the sky plane, we use the plane-of-sky projection of the two components ($\sigma_{R,\phi}\cdot\cos i$ and $\sigma_{z}\cdot \sin i$) and take their sum in quadrature:
\begin{equation}
    \sigma_{\rm{pm}}(r) = \sqrt{\sigma_{R,\phi}^2(r) \cos^2 i + \sigma_{z}^2(r) \sin^2 i}  \label{eq:model-2d-disp}
\end{equation}
\noindent where $i=37.4^\circ$ is the inclination of the LMC disk with respect to our line-of-sight adopted in B22. Note that V18 do not provide a single value for the inclination but rather a range of $32^\circ-35^\circ$. We have verified that this small range of inclinations does not affect our results. Further, using the projected dispersion in the observed PM (instead of the dispersion in an LMC-centric coordinate system) ensures there are no inconsistencies due to different choices of coordinate systems in V18 and C22.

$Q_{\rm{disp}}$ is calculated using eq. (\ref{eq:model-2d-disp}), and we refer to the scenario described in this subsection as the \enquote{ \modpm\ scenario }.

\subsubsection{Using a 3D Kinematic Model for the Field Star Velocities}
\label{sec:fieldmodel}
The C22 model provides the full 3D velocities for the field stars at the GC's location. Since the B22 catalogue also reports the GC's own 3D velocity, we calculate the full 3D velocity difference (instead of just the PM difference), and call it $\Delta_{\rm{3D}}$. As mentioned in section \ref{subsec:choi_model}, all GC and field velocities are converted to PMs and RVs before computing differences and statistical metrics. A conversion factor of $1$ mas yr$^{-1}$ $= 237.21$ km/s at 50 kpc \citep{Pietrzy_ski_2013} is used to convert RVs in km s$^{-1}$ to mas yr$^{-1}$ to match the units of PM. Similar to section \ref{subsec:model-error-sigma}, the error $\epsilon_{\rm{3D}}$ is calculated using Gaussian propagation from the errors in the GC's PM and RV. The errors in the field velocities calculated from the C22 model are taken to be $0$. 

Velocity dispersion $\sigma_{\rm{3D}}$ is computed using the red giant model in \citet[][hereafter V18]{Vasiliev_2018}. For the full 3D velocity, we take the sum of $\sigma_{R,\phi}$ and $\sigma_{z}$ in quadrature without the inclination factor. 
\begin{equation}
     \sigma_{\rm{3D}}(r) = \sqrt{\sigma_{R,\phi}^2(r) + \sigma_{z}^2(r)}  \label{eq:model-3d-disp}    
\end{equation}
$Q_{\rm{err}}$ and $Q_{\rm{disp}}$ are calculated (similar to \S~\ref{sec:statmetrics}) as follows: 
\begin{align}
    Q_{\rm{err}} &= \frac{\Delta_{\rm{3D}}}{\epsilon_{\rm{3D}}} \\ 
    Q_{\rm{disp}} &= \frac{\Delta_{\rm{3D}}}{\sigma_{\rm{3D}}}
\end{align}
The scenario described in this subsection is henceforth referred to as the \enquote{ \modrv\ scenario }.

To summarize our statistical framework, we have $Q_{\rm{err}}$ and $Q_{\rm{disp}}$ values under three different scenarios - 
\begin{enumerate}[a)]
    \item {\em \data\ scenario:} Using the PM difference between the GC (taken from B22) and the field RC stars (computed using \textit{Gaia} DR3). The velocity dispersions of the field stars are also computed using the \textit{Gaia} DR3 measurements.
    \item {\em \modpm\ scenario:} Using the PM difference between the GC (taken from B22) and the field RC stars (computed using the C22 model). The in-plane velocity dispersions of the field stars is computed from the red giant model of V18. 
    \item {\em \modrv\ scenario:} Using the 3D velocity differences between GC (taken from B22) and the field RC stars (computed using C22 model). The 3-D velocity dispersions of the field stars is computed from the red giant model of V18.
\end{enumerate}
With the model-driven field velocities, we use both \modpm\ and \modrv\ to ensure that the LoS velocity differences alone do not significantly increase the 3D velocity difference and bias our set of kinematic outliers.

\subsection{Line-of-Sight Effects in the Field Star Sample and Statistical Metrics}
\label{subsec:los-effects}
 Our selection criteria as detailed in the \S~\ref{sec:data} pick out RC field stars in the entire line-of-sight (LoS) across the LMC disk, and not specifically stars around a GC in 3-D. As such, the mean velocity components and their dispersions we measure in the following sections will include all LoS field stars instead of just the field stars immediately surrounding the GC in 3-D. Current distance precision within the LMC disk is not sufficient to accurately bin the RC stars along the LoS. Moreover, current distance estimates to the GCs themselves have uncertainties comparable to the LMC disk thickness (Figure~\ref{fig:veldiff}). 

The impact of the aforementioned LoS effect on our statistical metrics can be assessed as follows: the inferred velocity dispersion of the field star sample will be higher if all field stars are used instead of just the field stars in the immediate vicinity of a GC. This means that computing the significance of the velocity difference with respect to this overestimated dispersion will yield more conservative results.

Further, even though the LMC has a significant LoS depth, specially closer to the central regions \citep{Subramaniam2008}, the bulk of the stars are expected to reside within a scale height from the midplane \citep{van_der_Marel_2001, van_der_Marel_2002} of the LMC disk. Most studies attempting to model the kinematics of the LMC assume all stars of a chosen population as belonging to this main disk (see for example \citealt{Choi_2022, jimenez-arranz2023, VMC2025, Dhanush2024}). Therefore, we also assume that the mean PMs we compute reflect the local PM field of the disk. In such a scenario, if a GC has significantly different velocities, it means that either the GC is a part of the main disk and has anomalous kinematics, or that the difference is due to the GC being significantly above/below the main disk. Both of these possibilities are interesting, as in the former case the GC could have been accreted from an external environment, and in the latter case the GC could constitute the LMC's halo.

\section{A Catalogue of Outlying LMC Globular Clusters} \label{sec:results}

We select GCs that have $Q_{\rm{err}} > 3$ (velocity difference between the GC and field stars is inconsistent with $0$ at $>3\sigma$ level under measurement errors) and $Q_{\rm{disp}} > 1$ (velocity difference between the GC and field stars is greater than the velocity dispersion of the field stars). Under standard Gaussian statistics in a sample size of 30, the probability that any GC satisfies these two criteria simultaneously purely due to random fluctuations  is $<0.003$. Further, Tables~\ref{table:Mdata}~and~\ref{table:Mmodel} provide the values of $Q_{\rm{err}}$ and $Q_{\rm{disp}}$ for each cluster in each of the three methods; and the reader can choose their own threshold criteria.

The sample of GCs satisfying the above criteria are different under the three scenarios considered (\data, \modpm\ and \modrv). Importantly, small differences between a GC RV and the model-predicted RV for its surroundings, coupled with low uncertainties on the RV of the GC can increase $Q_{\rm{err}}$ and $Q_{\rm{disp}}$ under \modrv. As a result, the criteria defined in the above paragraph ($Q_{\rm{err}} > 3$ and $Q_{\rm{disp}} > 1$) flag about a third of the GC sample as kinematic outliers under the scenario \modrv.

Analyzing the statistical significance of the 2D velocity differences gives us more conservative results, in the sense that less number of GCs are flagged as outliers. \modpm\ flags 6 GCs as kinematic outliers, out of which 5 are flagged in \data\ as well. All of these 6 GCs under \modpm\ were also flagged as outliers under \modrv. In general, it would be expected for the \data\ or \modpm\ scenarios to give more conservative results as compared to \modrv, since \modrv\ picks out the GCs that may not have high PM differences but have high RV differences with respect to their surroundings.

The full set of results - GC PMs and RVs, data-derived PMs for the surroundings, model-derived velocities for the surroundings, and the values of $Q_{\rm{err}}$ and $Q_{\rm{disp}}$ under \data, \modpm\ and \modrv\ scenarios for all the LMC GCs considered in this study are given in Tables~\ref{table:Mdata} and \ref{table:Mmodel}.

\begin{table*}\centering
\begin{adjustbox}{width=1.2\linewidth,right}
  \begin{tabular}{c c c c c c c c c}
    \hline \hline
    Name of & GC Proper & GC Proper & Field Proper & Field Proper & Proper Motion & 2D Velocity & & \\

    GC & Motion in RA & Motion in Dec & Motion in RA & Motion in Dec & Difference $\Delta_0 \pm \epsilon$ & Dispersion $\sigma_v$ & $Q_{\rm{err}}$ & $Q_{\rm{disp}}$  \\ 
    
     & [mas yr$^{-1}$] & [mas yr$^{-1}$] & [mas yr$^{-1}$] & [mas yr$^{-1}$] & [mas yr$^{-1}$] & [mas yr$^{-1}$] &  &  \\
    
    \hline
    
NGC 1644 & -0.151 $\pm$ 0.123 & -0.679 $\pm$ 0.131 & -0.108 $\pm$ 0.027 & -0.633 $\pm$ 0.028 & 0.063 $\pm$ 0.13 & 0.12 & 0.485 & 0.528 \\
NGC 1651 & 0.082 $\pm$ 0.178 & -0.7 $\pm$ 0.158 & 0.094 $\pm$ 0.025 & -0.638 $\pm$ 0.025 & 0.063 $\pm$ 0.161 & 0.155 & 0.392 & 0.407 \\
NGC 1652 & -0.076 $\pm$ 0.325 & -1.042 $\pm$ 0.339 & 0.015 $\pm$ 0.025 & -0.648 $\pm$ 0.025 & 0.405 $\pm$ 0.339 & 0.146 & 1.194 & 2.765 \\
NGC 1756 & 0.0 $\pm$ 0.01 & -0.424 $\pm$ 0.01 & 0.058 $\pm$ 0.025 & -0.437 $\pm$ 0.025 & 0.059 $\pm$ 0.027 & 0.241 & 2.24 & 0.247 \\
NGC 1783 & -0.233 $\pm$ 0.043 & -0.394 $\pm$ 0.05 & -0.165 $\pm$ 0.025 & -0.382 $\pm$ 0.025 & 0.069 $\pm$ 0.05 & 0.152 & 1.382 & 0.453 \\
NGC 1786 & 0.094 $\pm$ 0.098 & -0.321 $\pm$ 0.159 & -0.061 $\pm$ 0.024 & -0.347 $\pm$ 0.025 & 0.157 $\pm$ 0.103 & 0.19 & 1.522 & 0.825 \\
NGC 1806 & -0.002 $\pm$ 0.064 & -0.418 $\pm$ 0.067 & -0.051 $\pm$ 0.024 & -0.326 $\pm$ 0.024 & 0.104 $\pm$ 0.071 & 0.219 & 1.477 & 0.476 \\
NGC 1831 & -0.178 $\pm$ 0.045 & -0.376 $\pm$ 0.061 & -0.197 $\pm$ 0.025 & -0.295 $\pm$ 0.025 & 0.084 $\pm$ 0.065 & 0.133 & 1.279 & 0.628 \\
NGC 1866 & -0.292 $\pm$ 0.047 & -0.182 $\pm$ 0.072 & -0.233 $\pm$ 0.025 & -0.193 $\pm$ 0.025 & 0.06 $\pm$ 0.054 & 0.149 & 1.114 & 0.404 \\
NGC 1868 & -0.24 $\pm$ 0.167 & -0.173 $\pm$ 0.257 & -0.197 $\pm$ 0.026 & -0.211 $\pm$ 0.026 & 0.057 $\pm$ 0.213 & 0.139 & 0.269 & 0.413 \\
NGC 1928 & 0.011 $\pm$ 0.01 & -0.07 $\pm$ 0.01 & -0.022 $\pm$ 0.027 & 0.009 $\pm$ 0.027 & 0.085 $\pm$ 0.029 & 0.453 & 2.954 & 0.188 \\
NGC 1939 & 0.113 $\pm$ 0.01 & -0.075 $\pm$ 0.01 & 0.043 $\pm$ 0.025 & 0.03 $\pm$ 0.025 & 0.126 $\pm$ 0.027 & 0.388 & 4.65 & 0.325 \\
NGC 1987 & 0.164 $\pm$ 0.009 & 0.027 $\pm$ 0.009 & 0.138 $\pm$ 0.024 & 0.079 $\pm$ 0.024 & 0.058 $\pm$ 0.026 & 0.287 & 2.235 & 0.202 \\
NGC 2108 & -0.215 $\pm$ 0.01 & 0.365 $\pm$ 0.01 & -0.208 $\pm$ 0.026 & 0.391 $\pm$ 0.026 & 0.027 $\pm$ 0.028 & 0.226 & 0.964 & 0.118 \\
NGC 2159 & -0.316 $\pm$ 0.075 & 0.509 $\pm$ 0.031 & -0.264 $\pm$ 0.025 & 0.515 $\pm$ 0.025 & 0.052 $\pm$ 0.079 & 0.155 & 0.662 & 0.336 \\
NGC 2162 & -0.415 $\pm$ 0.084 & 0.472 $\pm$ 0.076 & -0.33 $\pm$ 0.027 & 0.374 $\pm$ 0.027 & 0.13 $\pm$ 0.084 & 0.119 & 1.545 & 1.093 \\
NGC 2173 & 0.081 $\pm$ 0.043 & 0.501 $\pm$ 0.031 & 0.012 $\pm$ 0.025 & 0.493 $\pm$ 0.025 & 0.069 $\pm$ 0.05 & 0.137 & 1.391 & 0.504 \\
NGC 2190 & 0.028 $\pm$ 0.052 & 0.506 $\pm$ 0.056 & 0.014 $\pm$ 0.027 & 0.481 $\pm$ 0.027 & 0.029 $\pm$ 0.061 & 0.149 & 0.467 & 0.192 \\
NGC 2209 & 0.029 $\pm$ 0.154 & 0.582 $\pm$ 0.171 & -0.025 $\pm$ 0.026 & 0.573 $\pm$ 0.027 & 0.055 $\pm$ 0.157 & 0.097 & 0.348 & 0.56 \\
Hodge 4 & -0.227 $\pm$ 0.08 & 0.004 $\pm$ 0.09 & -0.289 $\pm$ 0.025 & 0.078 $\pm$ 0.025 & 0.096 $\pm$ 0.09 & 0.168 & 1.071 & 0.573 \\
NGC 1835 & 0.135 $\pm$ 0.013 & -0.38 $\pm$ 0.024 & 0.055 $\pm$ 0.026 & -0.255 $\pm$ 0.026 & 0.149 $\pm$ 0.033 & 0.323 & 4.448 & 0.46 \\
NGC 1898 & 0.137 $\pm$ 0.017 & -0.092 $\pm$ 0.02 & 0.045 $\pm$ 0.026 & -0.071 $\pm$ 0.026 & 0.094 $\pm$ 0.031 & 0.444 & 3.045 & 0.212 \\
NGC 1916 & -0.031 $\pm$ 0.055 & 0.119 $\pm$ 0.083 & -0.022 $\pm$ 0.027 & -0.027 $\pm$ 0.027 & 0.146 $\pm$ 0.087 & 0.464 & 1.678 & 0.315 \\
NGC 2005 & 0.144 $\pm$ 0.044 & 0.301 $\pm$ 0.063 & -0.059 $\pm$ 0.025 & 0.166 $\pm$ 0.025 & 0.243 $\pm$ 0.057 & 0.424 & 4.304 & 0.574 \\
NGC 2019 & 0.044 $\pm$ 0.034 & 0.038 $\pm$ 0.075 & -0.039 $\pm$ 0.026 & 0.194 $\pm$ 0.026 & 0.177 $\pm$ 0.073 & 0.42 & 2.43 & 0.421 \\ \hline
NGC 1818 & -0.333 $\pm$ 0.032 & -0.249 $\pm$ 0.048 & -0.167 $\pm$ 0.025 & -0.321 $\pm$ 0.025 & 0.181 $\pm$ 0.043 & 0.177 & 4.217 & 1.019 \\
NGC 1978 & -0.074 $\pm$ 0.048 & 0.026 $\pm$ 0.058 & -0.277 $\pm$ 0.025 & 0.085 $\pm$ 0.025 & 0.211 $\pm$ 0.055 & 0.185 & 3.854 & 1.14 \\
NGC 2210 & -0.309 $\pm$ 0.066 & 0.971 $\pm$ 0.025 & -0.288 $\pm$ 0.025 & 0.671 $\pm$ 0.025 & 0.301 $\pm$ 0.035 & 0.127 & 8.482 & 2.37 \\
NGC 2231 & -0.158 $\pm$ 0.046 & 0.77 $\pm$ 0.114 & -0.366 $\pm$ 0.025 & 0.706 $\pm$ 0.025 & 0.217 $\pm$ 0.061 & 0.114 & 3.585 & 1.907 \\
Hodge 11 & -0.393 $\pm$ 0.034 & 0.614 $\pm$ 0.049 & -0.267 $\pm$ 0.025 & 0.685 $\pm$ 0.026 & 0.145 $\pm$ 0.046 & 0.14 & 3.165 & 1.034 \\
     \hline 
  \end{tabular}
  \end{adjustbox}
  \caption{Comparison of PMs of the GCs (taken from \citealt{bennet2022kinematic}) against the average PMs of the field stars (computed from \textit{Gaia} DR3 data using a likelihood maximization scheme, \citealt{Vasiliev_2019a}). All PMs are in the LMC center-of-mass frame from C22 (\pmra$_{\mathrm{,COM}}=1.859$, \pmdec$_{\mathrm{,COM}}=0.375$). The last five GCs have a statistically significantly different PM from the field stars, with $Q_{\rm{err}} > 3$ (velocity difference is at least $3\sigma$ with respect to measurement uncertainties) and $Q_{\rm{disp}} > 1$ (velocity difference is greater than the velocity dispersion). For each GC named in the first column, we provide its PM components (columns 2 and 3), as well as the average PM components of its field stars (columns 4 and 5). The PM difference between the GC and field stars and the error in that difference is given in column 6. PM dispersion of the field stars is given in column 7. Our statistical metrics ($Q_{err}$ and $Q_{disp}$) for evaluating the significance of the PM difference between the GC and its field stars are given in columns 8 and 9. Derivation of PMs and dispersions of the surounding stars corresponding to each GC has been elucidated in section \ref{sec:fielddata}.}
  \label{table:Mdata}

\end{table*}

\begin{table*}\centering
\begin{adjustbox}{width=1.2\linewidth,right}
  \begin{tabular}{c c c c c c c c c c}
    \hline \hline
Name of & GC Radial & Field Model & Field Model & Field Model & Field Model & 2D Model & 2D Model & 3D Model & 3D Model \\ 
GC & Velocity & PM in RA & PM in Dec & RV (km/s) & 3D Dispersion &  $Q_{\rm{err}}$ & $Q_{\rm{disp}}$ & $Q_{\rm{err}}$ & $Q_{\rm{disp}}$ \\
 & [km s$^{-1}$] & [mas yr$^{-1}$] & [mas yr$^{-1}$] & [km s$^{-1}$] & [mas yr$^{-1}$] &   & & & \\

    \hline

NGC 1644  & -18.0$\pm$5.0 & -0.108 & -0.641 & -6.755 &  0.119  &  0.454  &  0.734  &  0.759  &  0.629 \\ 
NGC 1651  & -35.8$\pm$2.3 & 0.118 & -0.637 & -33.876 &  0.178  &  0.445  &  0.592  &  0.451  &  0.411 \\ 
NGC 1652  & 11.7$\pm$1.3 & 0.005 & -0.662 & -17.871 &  0.149  &  1.149  &  3.875  &  1.266  &  2.735 \\ 
NGC 1756  &  \ldots  & 0.042 & -0.425 & -11.857 &  \ldots  &  4.174  &  1.0  &  \ldots  &  \ldots \\ 
NGC 1783  & 27.6$\pm$9.9 & -0.189 & -0.39 & 18.217 &  0.135  &  1.019  &  0.486  &  1.388  &  0.437 \\
NGC 1786  & 16.0$\pm$5.0 & -0.068 & -0.372 & 3.504 &  0.164  &  1.616  &  1.511  &  1.768  &  1.083 \\ 
NGC 1831  & 16.0$\pm$5.0 & -0.224 & -0.279 & 26.363 &  0.119  &  1.85  &  1.381  &  2.133  &  0.983 \\ 
NGC 1866  & 34.4$\pm$0.4 & -0.252 & -0.17 & 30.657 &  0.14  &  0.842  &  0.446  &  0.958  &  0.32 \\ 
NGC 1868  & 19.0$\pm$3.0 & -0.246 & -0.169 & 34.618 &  0.128  &  0.034  &  0.08  &  2.791  &  0.521 \\
NGC 1898  & -54.0$\pm$5.0 & 0.04 & -0.075 & -5.615 &  0.24  &  5.748  &  0.579  &  11.113  &  0.944 \\ 
NGC 1916  & 14.0$\pm$5.0 & 0.018 & -0.048 & -2.527 &  0.245  &  2.145  &  1.008  &  2.473  &  0.766 \\
NGC 1928  & -14.4$\pm$12.8 & 0.017 & -0.011 & -2.255 &  0.24  &  5.882  &  0.347  &  2.152  &  0.326 \\ 
NGC 1939  & -5.2$\pm$7.4 & 0.05 & 0.0 & -6.513 &  0.25  &  9.807  &  0.555  &  9.697  &  0.393 \\ 
NGC 1987  &  \ldots  & 0.09 & 0.096 & -11.866 &  \ldots  &  11.199  &  1.0  &  \ldots  &  \ldots \\ 
NGC 2019  & 16.6$\pm$2.3 & 0.033 & 0.164 & -4.683 &  0.197  &  1.696  &  0.928  &  2.53  &  0.788 \\ 
NGC 2108  &  \ldots  & -0.085 & 0.342 & 8.711 &  \ldots  &  13.185  &  1.0  &  \ldots  &  \ldots \\
NGC 2162  & 38.0$\pm$3.5 & -0.351 & 0.38 & 62.256 &  0.131  &  1.428  &  1.282  &  2.577  &  1.158 \\
NGC 2173  & -26.6$\pm$0.7 & 0.075 & 0.522 & -18.468 &  0.149  &  0.695  &  0.221  &  2.353  &  0.275 \\ 
NGC 2209  & -9.0$\pm$3.0 & 0.015 & 0.616 & -16.272 &  0.129  &  0.217  &  0.427  &  0.367  &  0.37 \\ 
Hodge 4  & 46.8$\pm$1.9 & -0.292 & 0.092 & 42.318 &  0.168  &  1.268  &  0.959  &  1.304  &  0.665 \\ \hline
NGC 1806  & -39.0$\pm$5.0 & -0.054 & -0.321 & 2.956 &  0.173  &  1.663  &  0.931  &  5.298  &  1.209 \\ 
NGC 1835  & -76.0$\pm$5.0 & 0.044 & -0.26 & -8.385 &  0.22  &  7.3  &  0.973  &  15.373  &  1.468 \\ 
NGC 2159  & -211.0$\pm$30.0 & -0.196 & 0.548 & 18.157 &  0.196  &  1.754  &  0.925  &  7.749  &  4.983 \\ 
NGC 2190  & -4.0$\pm$3.0 & 0.07 & 0.517 & -24.092 &  0.088  &  0.827  &  0.735  &  3.605  &  1.08 \\
\hline
NGC 1818  & 47.0$\pm$4.0 & -0.169 & -0.307 & 17.739 &  0.15  &  5.092  &  1.723  &  7.218  &  1.423 \\ 
NGC 1978  & 28.0$\pm$2.5 & -0.245 & 0.077 & 31.198 &  0.16  &  3.656  &  1.642  &  3.677  &  1.121 \\ 
NGC 2005  & 6.0$\pm$5.0 & 0.01 & 0.134 & -1.417 &  0.205  &  3.803  &  1.491  &  3.878  &  1.058 \\ 
NGC 2210  & 79.0$\pm$5.0 & -0.227 & 0.696 & 19.974 &  0.152  &  9.414  &  2.807  &  14.13  &  2.497 \\ 
NGC 2231  & 13.6$\pm$1.4 & -0.343 & 0.715 & 41.859 &  0.143  &  3.523  &  2.004  &  4.855  &  1.592 \\ 
Hodge 11  & -18.9$\pm$1.0 & -0.197 & 0.728 & 14.801 &  0.15  &  5.909  &  2.247  &  8.218  &  1.788 \\

     \hline
  \end{tabular}
  \end{adjustbox}
  \caption{Comparison of the GC velocities (taken from B22) with the average field RC stars' velocities computed using the C22 kinematic model. All PMs and RVs are in the LMC center-of-mass frame from C22 (\pmra$_{\mathrm{,COM}}=1.859$, \pmdec$_{\mathrm{,COM}}=0.375$, RV$_{\mathrm{COM}}=264.05$ km s$^{-1}$). The last 6 GCs have a significantly different PM and 3D velocity as compared to the model predictions. The preceding 4 GCs have a significantly different 3D velocity as compared to the model prediction, but they do not have a significantly different PM. For each GC in column 1, we list its RV in column 2, followed by the field star model velocities and velocity dispersions in columns 3--6. The statistical metrics for evaluating the significance of the 2D and 3D velocity differences between a GC and its field stars are given in columns 7--10. Velocity dispersions are computed using the 3D velocities, and are quoted in mas yr$^{-1}$ for easy comparison (at LMC distance, $1$ mas yr$^{-1}$ $= 237.2$ km s$^{-1}$). NGC 1756, 1987 and 2108 did not have RV data or precise LoS distances calculated in B22, so their entries are empty in some of the columns. Derivation of PMs and RVs of a GC's field stars has been elucidated in section \ref{sec:fieldmodel}. 
  }
  \label{table:Mmodel}
\end{table*}

\begin{figure*}
  \centering
\includegraphics[width=0.49\textwidth]{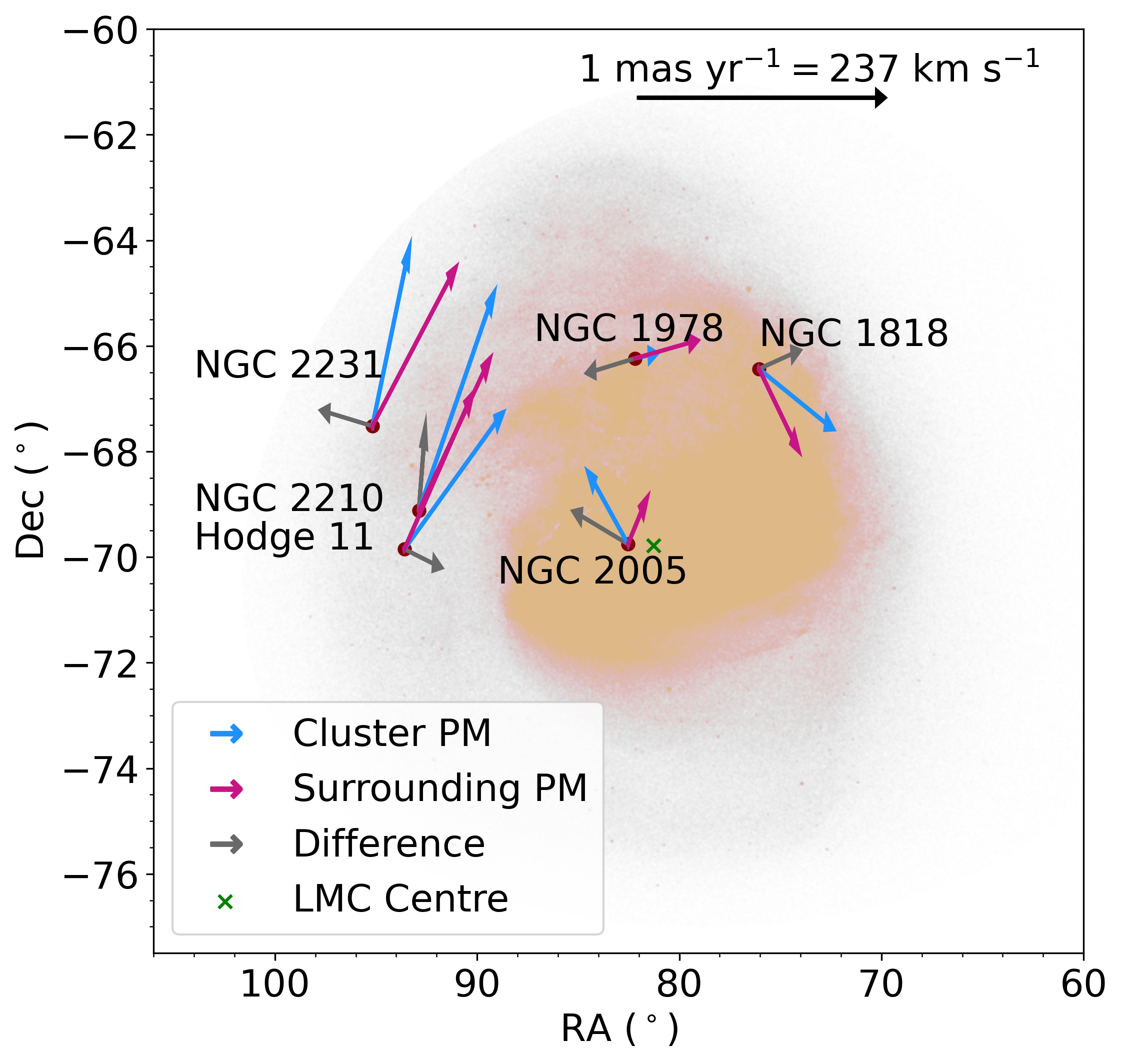}
\includegraphics[width=0.49\textwidth]{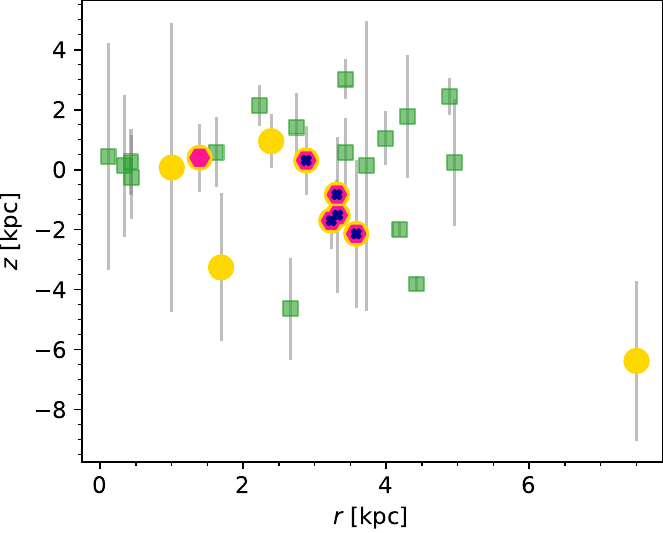}
    \caption{{\em Left panel:} PM difference vectors (in grey) between the GC (in blue) and their field RC stars (in magenta). Only the 6 GCs having a statistically significant difference with their field stars under both \textit{Gaia} DR3 data (\data) and the RC kinematic model (\modpm) are shown (see Table \ref{table:Mmodel}). These GCs are overlayed on a background of LMC stars obtained from \textit{Gaia} DR3. The surrounding field PMs clearly show the expected rotational pattern. While the directions of the velocity differences do not show any trend, 5 of the 6 GCs are located at similar distances ($\sim 3-4$ kpc) from the LMC photometric center (marked with a green cross). {\em Right panel:} Distribution of the GCs in the $r$-$z$ space of the LMC-centric coordinate system. Green squares are all the non-outlier GCs. Outlying GCs under \modrv\ , \modpm\ , and \data\ , are marked with yellow circles, pink hexagons and blue crosses respectively. Grey errorbars denote the $1-\sigma$ uncertainty in $z$. Uncertainties in $r$ are negligible compared to the uncertainty in $z$. Consistent with the finding in the {\em left panel}, the high difference GCs under all three scenarios (marked with yellow, pink and blue) are clustered in the $r$ - $z$ space. The 2D Kolmogorov-Smirnov test returns a $p$-value $\approx 0.01$ for the hypothesis that both the set of outlier GCs and the set of remaining GCs were drawn from the same underlying distribution, indicating that their clustering is statistically significant.}
\label{figure:dirs}
\end{figure*}

\begin{figure}
\centering
\includegraphics[width=\columnwidth]{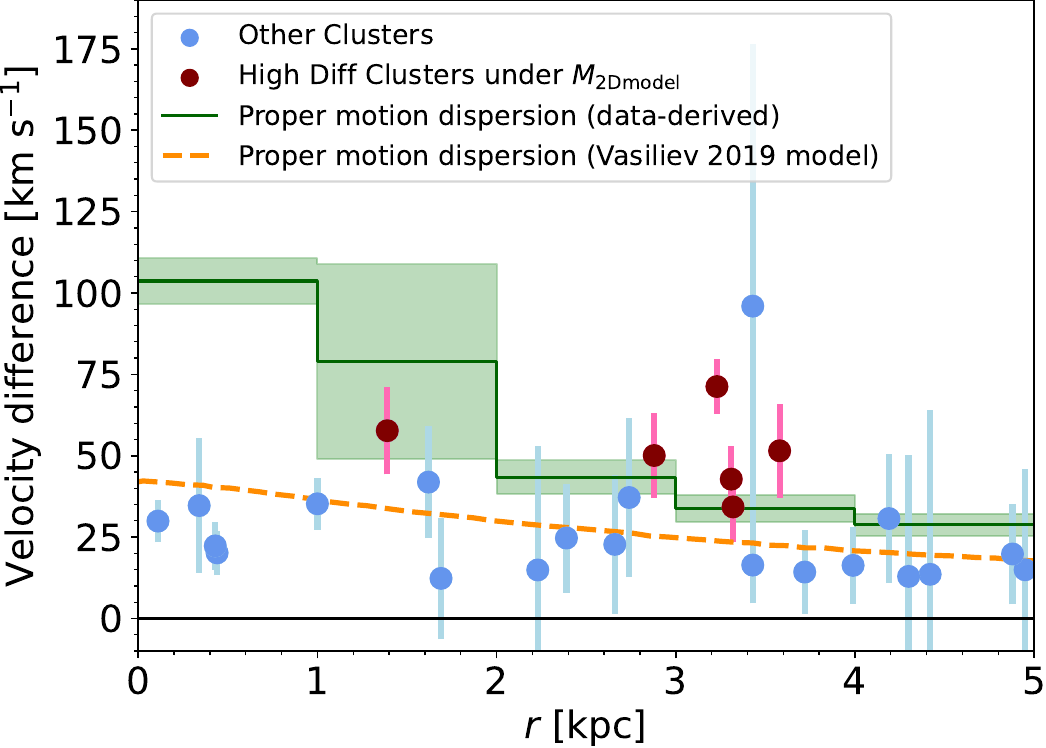}
\caption{GC velocity differences are compared with the velocity dispersion of the field stars, as a function of the radial coordinate (R). The velocity dispersions are obtained in two ways - using \textit{Gaia} DR3 data (green solid line) and using the RGB model of \cite{Vasiliev_2018} (orange dashed line). The 6 high difference GCs identified in Table \ref{table:Mmodel} are marked in pink and the remaining GCs are marked in blue. $1-\sigma$  errorbars on the velocity differences are also shown. The green band denotes the standard deviation of the velocity dispersion values in each distance bin (1--2 kpc, 2--3 kpc and so on), with the average velocity dispersion for that distance bin marked with the solid green line.}
\label{fig:veldiff}
\end{figure}

The PM vectors of the 6 GCs identified as outliers in Table \ref{table:Mdata} are plotted in Figure \ref{figure:dirs} {\em left panel}. The PM vectors of the field stars as well as the PM difference vectors between the GC and the surroundings are also shown. We do not find any trend in the directions of the PM difference vectors across the 6 GCs. However, surprisingly, 5 out of 6 outlying GCs are located at a similar radial distance (3--4 kpc) from the LMC's photometric center. Interestingly, \cite{Olsen2011} found a population of stars potentially accreted from the SMC also at a radius of $\sim 3$ kpc. We note that this is also approximately the radius where the rotation curve changes from rising to flat and could potentially contribute to artificial velocity residuals when comparing against the model (\modpm). However, this is unlikely to be the case since the GCs around 3--4 kpc are outliers selected by both \data\ and \modpm\ scenarios.

The distribution of the LMC GC sample in the $r-z$ space is visualized in Figure \ref{figure:dirs} {\em right panel}. Uncertainties in $z$ are taken to be the same as  the distance uncertainties given in Table 1 of B22. This is justified since the LMC is approximately face-on. Uncertainties in $r$ are negligible as compared to the uncertainty in $z$. The 5 outliers under the scenario \data\ are found to cluster in the $r - z$ space, consistent with Figure \ref{figure:dirs} {\em left panel}. We perform the 2D Kolmogorov–Smirnov (KS) test \citep{Peacock_1983, Fasano_1987} on the full distribution of $(r,z)$ v/s the $(r,z)$ of the 5 kinematic outliers flagged under all three scenarios. We use the public code \textsc{ndtest}\footnote{Written by Zhaozhou Li, \url{https://github.com/syrte/ndtest}}. The $p$ value for the two sets being drawn from the same underlying distribution is $0.046$, indicating that the kinematically outlying set has a statistically different distribution of $(r,z)$ values as compared the full population of GCs. This further drops to $p=0.011$ if we compare the outlier cluster set against the non-outlier cluster set only (instead of the full population).

It is also reasonable to expect that the outskirts of the LMC would be more prone to kinematic perturbations due to tidal interactions with other galaxies including the SMC and MW. That could cause GCs at large radii to naturally have large velocity differences. However, from Tables~\ref{table:Mdata}~and~\ref{table:Mmodel} we see that there is no significant correlation or increase of either the PM difference or the values of $Q_{\rm{err}}$ and $Q_{\rm{disp}}$ with the radius. As such, a GC being at a larger radius does not seem to influence its velocity difference significantly.

We emphasize the necessity of using both the metrics $Q_{\rm{err}}$ and $Q_{\rm{disp}}$ for selecting a reliable set of high-difference GCs. High velocity differences between the GCs and their field stars are found in the LMC's central regions. However, the LMC's central region has intrinsically high PM dispersions, and hence these GCs cannot be called outlying unless the velocity differences exceed the local velocity dispersion. Similarly, GCs at large radii ($r \gtrsim 5$ kpc) have high velocity differences, but they are located in sparser regions of LMC resulting in higher uncertainties on the mean PMs of field stars. Hence, these GCs cannot be called outlying unless the velocity difference is statistically significant over the PM uncertainties. It is necessary to use both the metrics, $Q_{\rm{err}}$ and $Q_{\rm{disp}}$, for carefully evaluating the significance of the GC velocity differences. 

As a consistency check of our PM dispersion calculations of the field stars with Gaia DR3 data (section \ref{subsec:selection-radii}), we compare the data-derived RC dispersions with the red giant based dispersion model of V18. This comparison is shown in Figure \ref{fig:veldiff}. We also plot the PM difference (along with errors) between the GC and its surroundings for the entire sample of GCs utilized in this work. The mean RC PM dispersions calculated with Gaia DR3 data are systematically higher as compared to the red giant model based dispersion curve. Although, for $r > 2$ kpc, this discrepancy is within the $3-\sigma$ uncertainty of the data-derived dispersions. Note that the area over which we compute the RC PM dispersion (a circle of radius $\approx 0.15^\circ$) is larger than that used in V18 (a square of side $0.2^\circ$). Further, while the V18 kinematic model is tailored to the mean PM of the LMC disk sp, our selection criteria of all RC stars along the line-of-sight would also include stars that might be significantly above or below the main disk, driving up the observed dispersion. Hence, there is a higher physical spread of velocities in our selected regions as compared to the spatial bins used in V18, which can partly account for the differences in the two dispersions for $r > 2$ kpc. 

The mean discrepancy between the two dispersions is much larger in the LMC's inner 2 kpc, wherein the data derived dispersions are a factor of 2-3 larger than the model derived dispersions. The large discrepancy can be attributed to the high stellar crowding in the LMC's inner regions \citep{Rathore_2025}. Indeed, the $1-\sigma$ errorbar on the data derived dispersion is $\approx 25$ km s$^{-1}$, which would make the data derived dispersions consistent with the model derived ones at the $2-\sigma$ level. The data derived dispersions in the LMC's inner 1 kpc are likely not reliable due to severe crowding. This further emphasizes the need to compare the GC velocity differences with both the data derived and the model derived PM differences, for identifying a sample of GCs that are truly kinematically outlying. 

\section{Discussion} \label{sec:discussion}

We have developed a statistical framework to select LMC GCs that are kinematically distinct from their field LMC disk stars. We find five GCs (NGC 1818, NGC 1978, NGC 2210, NGC 2231 and Hodge 11) that have velocities that are significantly different from both data-derived and model-derived 2D and 3D velocities of the field stars. One more GC (NGC 2005) has a significantly different velocity from only the model-derived 2D and 3D velocities of the surroundings. Four others (NGC 1806, NGC 1835, NGC 2159 and NGC 2190) have a significant differences only in the line-of-sight component of their velocity, and not their PM component. Next: (i) we place our results in the context of some previous studies on the LMC GCs; (ii) examine the dependence of the velocity differences on the age of GCs; (iii) investigate the reliability of the GC sample as a tracer of the LMC's mass; (iv) assess the limitations of our analysis and prospects for the future.

\subsection{Previous Investigations of Our Outlying GCs}

Utilizing the full 3D velocity information (\modrv) yields $10$ kinematic outliers. Of these, $4$ are not outliers in the PM space, implying that they are outliers in the LoS velocity only. Of these four, NGC 2159 was also analyzed by \cite{bennet2022kinematic} as a potential outlier based on its anomalous LoS velocity and discrepancy with the LMC's average rotational field. 

We note that three of the four GCs that are outliers only in the LoS velocity (NGC 1835, NGC 2159, NGC 2190) have RV estimates based on only a few ($\sim 5$) member stars \citep{Baird1986, Olszewski1991, Sharma2010}. Hence, the RV measurements for these GCs may not be very robust, and additional spectroscopic follow up is necessary. In this subsection, we only focus on the remaining $6$ GCs, which are kinematic outliers in the PM space as well. We investigate previous works to find further evidence of anomalous behavior of these GCs.

NGC 1978 is a globular cluster with highly elliptical isodensity curves ($\epsilon \approx 0.30$, \citealt{Mucciarelli_2007}), quite unlike most MW GCs (mean ellipticity $\approx 0.07$, and typical values between 0 and 0.15, \citealt{goodwin1997ellipticities, Freour2026}) and lies at the upper end of the ellipticity range of other LMC GCs (0.05--0.30, with a mean $\approx 0.14$, \citealt{goodwin1997ellipticities}). In general, LMC GCs have higher ellipticities as compared to the MW GCs. This is due to a weaker tidal field of the LMC compared to the MW, where in the former case the original triaxial structure of the GCs is preserved (\citealt{goodwin1997ellipticities} and references therein). However, the ellipticity of NGC 1978 is abnormally high. \cite{ferraro2006iron} showed that such high ellipticities {\em cannot} be achieved with mergers of two GCs or even two progenitor molecular clouds, which makes the high ellipticity of NGC 1978 quite intriguing.

The peculiar shape of NGC 1978, combined with the large PM difference with the surroundings, suggests that the GC might have been born in a different environment. We propose that the anomalous kinematics and extra high ellipticity of NGC 1978 makes it a promising candidate of having been originally formed in a dwarf galaxy with a lesser tidal field strength as compared to the LMC. If the GC was accreted into the LMC much later, the LMC's tidal forces would not have enough time to alter the shape of the GC.

In a recent study by \cite{mucciarelli2021relic}, NGC 2005 was found to have distinct chemical abundance patterns compared to other LMC GCs of similar metallicity, suggesting it originated in a different environment. Using chemical evolution models, they concluded that NGC 2005's progenitor was likely a low-mass dwarf spheroidal galaxy with inefficient star formation. NGC 2005 ending up as a kinematic outlier in our framework lends further credence to this GC having an external origin. However, note that \cite{Piatti_2023} refutes the findings of \cite{mucciarelli2021relic}.

B22 found that the speed of NGC 2210 relative to the LMC is likely larger than the LMC escape speed (see also Figure \ref{fig:E-Lz}). NGC 2210 is an old GC (age $>8$ Gyr, \citealt{olszewski1996old}), which shows multiple horizontal branches in its CMD \citep{Gilligan_2019, gilligan2020exploring}. Multiple populations are typically seen in cores of accreted dwarf galaxies as shown in \cite{Helmi_2008}, and such cores are likely progenitors of several MW GCs \citep{Massari_2019, Joel2021, Belokurov_2024}. Hence, it is likely that NGC 2210 is not a LMC GC, and instead is a MW GC that happens to be near the LMC.

Apart from NGC 2210, \cite{Gilligan_2019} observed that Hodge 11 is the only other LMC GC with multiple horizontal branches. Further, both NGC 2210 and Hodge 11 have the most heavily populated blue-straggler branch among the very old GCs (age $> 12$ Gyr) studied by \cite{Wagner_Kaiser_2017}. However, the relative speed of Hodge 11 with respect to the LMC suggests that it is bound to the LMC, unlike NGC 2210. Further investigations of Hodge 11 are required, as its unique characteristics --- being kinematically peculiar within the LMC disk and comparable in age to the oldest, metal-poor GCs in the MW --- make it an ideal target to investigate the origin of the earliest GCs in LMC-mass galaxies.

\subsection{Dependence of GC Kinematic Differences on Age}
As mentioned previously, the LMC GC population has a mix of ages, including young GCs (age $< 0.5$ Gyr), intermediate age GCs (0.5 Gyr $\leq$ age $\leq 4.0$ Gyr) and old GCs (age $\gtrsim 10$ Gyr). Our model and data-driven estimates of the field star PMs and RVs is derived using the RC population, which is dominated by ages of $\sim 2$ Gyr \citep{Girardi2016}. We therefore expect kinematic differences between a GC and its field stars to also arise due to the different epochs in the LMC's evolution at which the GC and field RC populations formed. The goal of this subsection is to understand to what extent are the kinematic differences of the GC population and the field stars correlate with the GC ages.

\begin{figure*}
\centering
\includegraphics[width=0.33\textwidth]{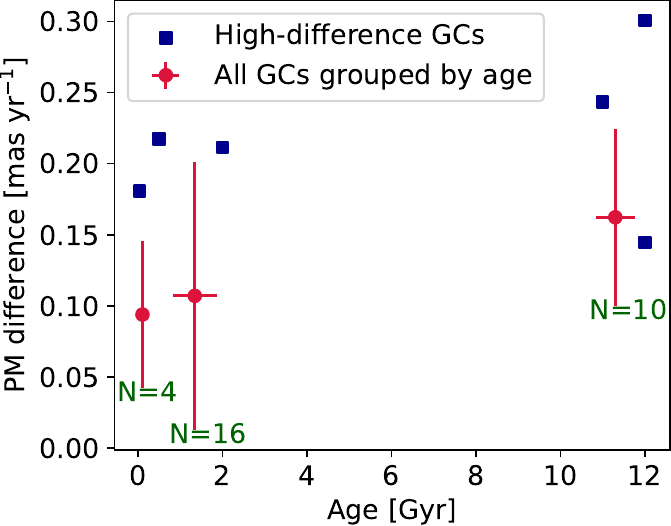}
\includegraphics[width=0.320\textwidth]{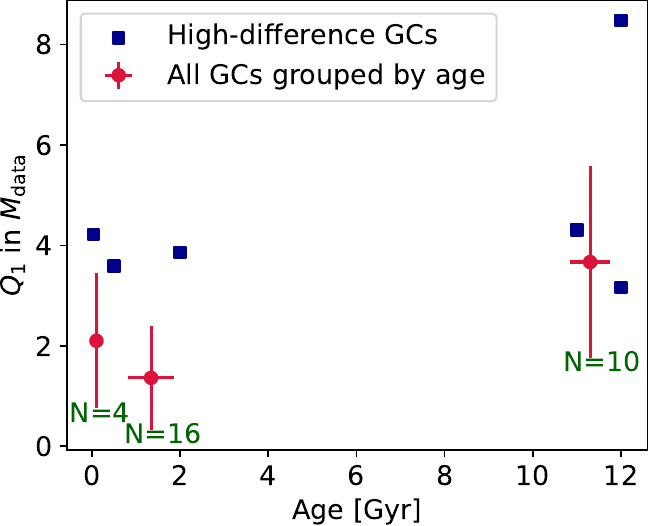} 
\includegraphics[width=0.333\textwidth]{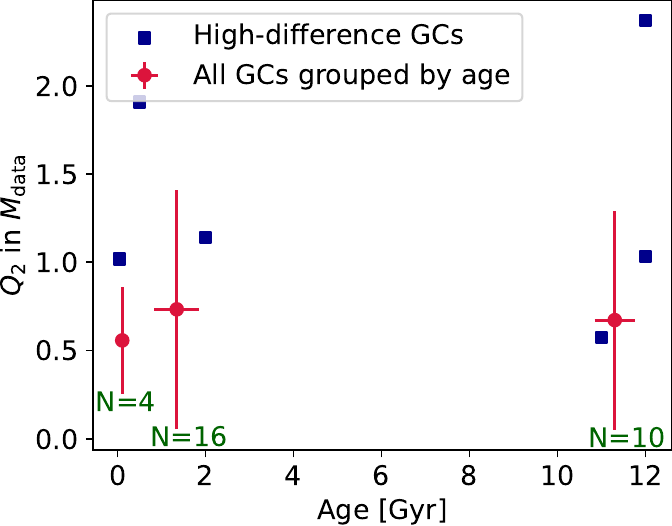}
\caption{Dependence of the kinematic differences between the LMC GC population and their field stars on the GC ages. The red points correspond to the entire GC sample analyzed in this work, categorized into young GCs (age $< 0.5$ Gyr, 4 in number), intermediate age GCs (0.5 Gyr $\leq$ age $\leq$ 4 Gyr, 16 in number) and old GCs (age $\gtrsim$ 10 Gyr, 10 in number). The blue squares denote the 6 outliers in the \modpm\ scenario. {\em Left panel} shows the PM difference between the GC and field stars as a function of GC age. When the entire GC sample is considerd, there is a slight mean trend of older GCs having higher kinematic differences. However, given the standard deviation of the PM difference in each age group, this trend is not statistically significant. Further, no correlation is evident between the PM difference and age of the 6 outlying GCs. There is no evidence of the statistical metrics $Q_{err}$ and $Q_{disp}$ having a statistically significant dependence on GC age ({\em middle panel} and {\em right panel} respectively).}
\label{fig:agevars}
\end{figure*}

A literature compilation of the ages of the GC sample analysed in this work are given in Table~\ref{tab:lmc-gc-ages}. We note that there are no GCs in the age range of 4--10 Gyr, which is well-known as an age gap in the LMC GC system \citep{Olszewski1991, agegap2004, agegap2022}. In Figure~\ref{fig:agevars} {\em left panel}, we plot: (i) the average and standard deviation of the PM difference of all the young, intermediate and old GCs considered in this work; and (ii) the PM difference as a function of age for the 6 outlying GCs identified under the \modpm\ scenario. There seems to be a slight mean trend of increasing PM difference as we go from young to intermediate to old GCs. However, given the size of the $1-\sigma$ scatter, this trend is not statistically significant. For each age category, we create 10000 mock values of the PM difference by sampling from a Gaussian with a mean at the mean PM difference observed for that age category, and standard deviation equal to the $1-\sigma$ scatter. We compute the Spearman correlation coefficient between the age and the PM difference for all 10000 realizations, and find its value to be $\rho = 0.42 \pm 0.54$, indicating that the correlation is consistent with 0. Further, no significant correlation is evident between the PM difference and ages of only the 6 outlying GCs (Spearman correlation coefficient = 0.23). In Figure~\ref{fig:agevars}, {\em middle} and {\em right} panels show the dependence of the statistical metrics $Q_{err}$ and $Q_{disp}$ on the GC ages. No correlation is evident between the value of these metrics and GC ages; neither for the outlier sample nor for the three age based GC categories. 

Hence, the kinematic differences of the outlying GC population and the field stars cannot be explained by GC age differences alone. 

Further, ages for the 5 GCs flagged as outliers under the \data\ scenario are: 2 Gyr for NGC 1978 \citep{Narloch_2022}, $>10$ Gyr for NGC 2210 and Hodge 11 \citep{Sagar_Ages,Johnson1999,Wagner_Kaiser_2017}, 40 Myr for NGC 1818 from \citep{Marino_2018}, $\sim$13 Gyr for NGC 2005 \citep{Olsen1998} and 1.6 Gyr for NGC 2231 \citep{piatti2014}. There is no evidence of clustering in ages of these kinematically outlying GCs. We note that NGC 1818 is a very young GC. Hence, it is possible that the kinematics of NGC 1818 is similar to its parent gas cloud, which could explain why its velocity is different from the field stars.

\subsection{Possibility of the Outlying GCs being Members of LMC Halo}
Another possible explanation for the anomalous kinematics of a subset of our GCs could be their 3D position with respect to the LMC's main disk (as noted in \S~\ref{subsec:los-effects}). If the GC, however, is located significantly above/below the main disk, it would naturally result in the GC having very different kinematics from the disk stars at the same projected location on the sky. Excitingly, such a GC could be a part of the LMC's stellar halo.

\cite{Piatti2019} analyzed the velocity component perpendicular to the disk for a subset of the LMC GCs. $3$ out of our 6 2D kinematic outliers reside in their sample as well --- NGC 2005, NGC 2210 and Hodge 11. While they classify NGC 2005 and NGC 2210 as disk GCs based on a low $|v_z|$, they interpret Hodge 11 as a halo GC. A more in-depth analyses, which combines precision distances, astrometry and chemical abundances, is required to rigorously ascertain if some of our kinematically outlying GCs constitute the LMC's halo. 

Cosmologically, LMC mass galaxies are expected to have a significant stellar halo population \citep{Tau2025, KadoFong2022}. However, the LMC's stellar halo has been elusive, partly because of the LMC being approximately face-on with a very extended disk. In particular, most of the stars at $R > 10$ kpc from the LMC's center are still consistent with an exponentially declining profile of the LMC's main disk \citep{Saha2010}. Such an extended disk makes it challenging to identify halo stars at large LMC-centric radii. An alternative approach is to look for populations above or below the LMC's disk in the LoS. However, directly identifying such stellar populations requires very precise distances that are currently not available. Recently, LoS velocity analysis of red giants in M33 by \cite{Gilbert2022} also provided evidence for a significant halo component based on kinematic differences with respect to the M33's HI disk. Thus a detailed analysis of the GC kinematics is a promising indirect step towards identifying the LMC's stellar halo.

\subsection{Implications for LMC Mass Estimate}

\begin{figure*}
\centering
\includegraphics[width=0.32\textwidth]{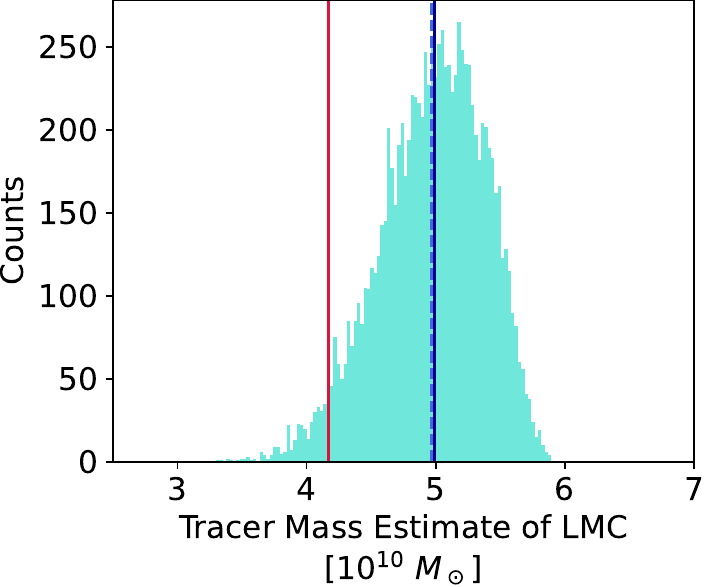}
\includegraphics[width=0.32\textwidth]{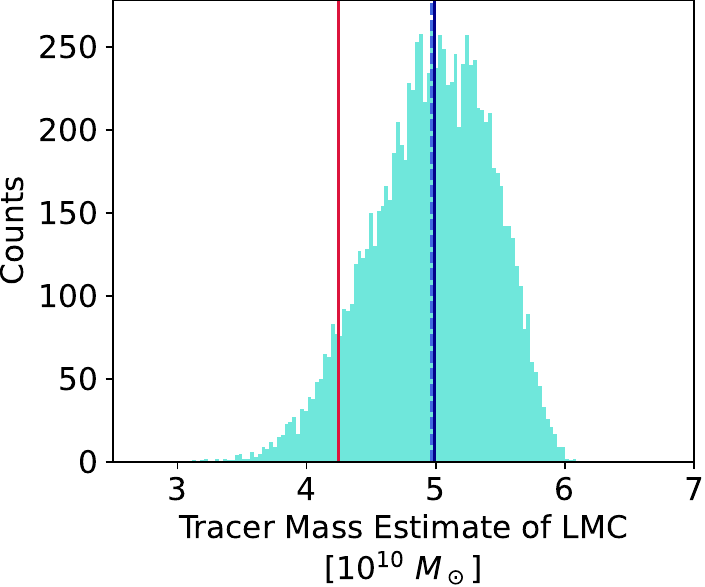} 
\includegraphics[width=0.32\textwidth]{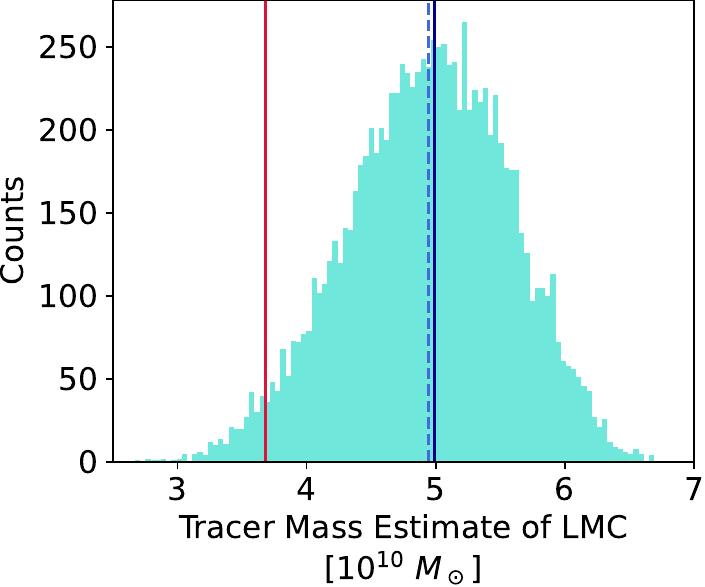}
\caption{The GC based LMC tracer mass estimate (TME) with the full GC sample (solid blue line) v/s the estimate obtained by removing the outlier GCs flagged by \data, \modpm\ and \modrv\ (solid red line in the {\em left}, {\em middle} and {\em right} panel respectively). The turquoise histogram depicts the TME distribution of the bootstrapped realizations obtained by randomly removing the same number of GCs as the oulying GCs and using only the remaining sample. The mean of the bootstrapped realizations (dashed blue line) is close to the full sample estimate as expected. 
We find that the kinematically outlying LMC GCs can bias the TME by as large as 30\%.}
\label{fig:tracermass}
\end{figure*}

Precise determination of the LMC's mass, particularly the dark matter (DM) contribution, is important for several purposes, like - identifying LMC's satellite galaxies \citep[e.g.][]{patel2020orbital}, constraining the LMC-SMC interaction history \citep[e.g.][]{Besla_2012, Kallivayalil_2013, patel2020orbital, Arranz2024}, placing the LMC-MW system in a cosmological context \citep[e.g.][]{Besla2018, Chamberlain2024, Arora2025}, and ultimately, using the LMC as a testbed for DM particle physics \citep[e.g.][]{Foote2023, Foote2026}. 

\cite{Watkins_2024} determined the mass profile of the LMC using the system of GCs in the B22 catalog. The tracer mass estimator (TME) used in their work is:
\begin{align} \label{eq:tme}
    M(<r_{\rm{max}}) = \frac{\alpha+\gamma-2\beta}{G(3-2\beta)} \cdot r_{\rm{max}}^{1-\alpha}\braket{v^2r^\alpha}
\end{align}
Here $G$ is the gravitational constant, $r$ and $v^2$ are the radial distance in the LMC-centric cylindrical coordinates $(r,\phi,z)$ and the total velocity modulus squared respectively. $\braket{}$ denotes the average taken over the full tracer sample. $\gamma$ is the power-law index for the radial distribution function of the GCs, $\beta$ is the velocity anisotropy in the GCs and $\alpha$ is the power-law index that can be assumed to describe the gravitational potential over the range of $r$ that is spanned by the cluster sample. 

\cite{Watkins_2024} assumed that the kinematics of their GC sample was an ideal tracer of the underlying gravitational potential. However, we have shown that a significant subset of the LMC GCs are kinematically outlying. Such outyling GCs may not be suitable tracers of the LMC's mass. In this subsection, we investigate the effect of including or removing the kinematically outlying GCs on the LMC's mass estimate.

We re-calculate the LMC's TME for three GC samples -- 
\begin{enumerate}[1)]
    \item Using the full sample used in our analysis
    \item The sample obtained after removing the 2D kinematic outliers 
    \item The sample obtained after removing the 3D kinematic outliers
\end{enumerate}
Note that our full sample does not correspond to the full sample in B22 catalogue, since we removed many GCs that do not have PM and RV measurements. We newly include three GCs (NGC 1466, NGC 1841, Reticulum) that were not included in our velocity difference analysis because their field star PMs could not be reliably estimated due to insufficient counts in the surrounding annuli. We also include NGC 2210 and NGC 2159 (not used by \citealt{Watkins_2024}), as these are part of the kinematic outlier sets.

Therefore, the values of $\alpha, \beta, \gamma$ obtained by \cite{Watkins_2024} would not be exactly the same for our sample. Re-evaluating these constants for our three samples is beyond the scope of this work. Moreover, evaluating the LMC's mass precisely using the GC sample is not the focus of this work. Instead, we seek to determine how much the TME changes when the outlying GCs are removed. We use the same values of $\alpha, \beta, \gamma$ as derived by \cite{Watkins_2024} for all three GC samples. 

The maximum radius in our sample (i.e. the radius for the enclosed mass estimate) is $R_{max} = 11.59$ kpc. The TME inferred enclosed mass within $R_{max}$ is $4.99 \times 10^{10} M_\odot$ for the full GC sample. Next, we remove the five GCs flagged as kinematic outliers in all three of \data, \modpm\ and \modrv, and recalculate the mass with the remaining GCs. The enclosed mass within $R_{max}$ is $4.17 \times 10^{10} M_\odot$. Hence, the enclosed mass changes by $16\%$ when the 2-D kinematic outliers are removed. 

We ascertain the statistical significance of the change in enclosed mass through a bootstrapping procedure. We create 1000 realizations of GC sets with 5 random GCs removed from the full sample in each case, and recalculate the TME with the new sets. The resulting mass distribution consists of estimates with a mean of $4.97 \times 10^{10} M_\odot$, almost equal to the mass estimate from the full sample, as expected. The estimate with the 5 outliers removed ($4.17 \times 10^{10} M_\odot$) is $\approx 2\sigma$ away from this mean, where $\sigma$ is the standard deviation of the mass distribution of bootstrapped realizations. The distribution of the bootstrapped realizations is approximately Gaussian, as shown in Figure \ref{fig:tracermass} {\em left panel}.

Next, we remove the 6 GCs flagged as outliers in \modpm\ and \modrv\ from the full sample, and obtain a TME inferred enclosed mass within $R_{max}$ of $4.24 \times 10^{10} M_\odot$. This new TME is different from the TME of the full sample by $15\%$. Repeating the bootstrapping analysis yields a statistical significance $\approx 1.65\sigma$ for this difference.  The distribution of the bootstrapped realizations is approximately Gaussian, as shown in Figure \ref{fig:tracermass} {\em middle panel}.

Finally, we remove the 10 GCs found to be outliers under \modrv\ from the full sample, and obtain a TME inferred enclosed mass within $R_{max}$ of $3.63 \times 10^{10} M_\odot$. This new TME is different from the TME of the full sample by $\approx 30\%$. Repeating the bootstrapping analysis yields a statistical significance of $\approx 2\sigma$ for this difference. The distribution of the bootstrapped realizations is approximately Gaussian, as shown in Figure \ref{fig:tracermass} {\em right
panel}.

To summarize, we find that including or excluding the kinematically outlying GCs can affect the LMC's enclosed mass by 15\% - 30\%, depending on the outlier sample used. This bias will directly impact the LMC's DM content, since the baryonic mass profile is well measured \citep{van_der_Marel_2002, Staveley-Smith2003, Olsen2007}. Further, the above bias is likely only a lower limit, since we have utilized the same value of $\alpha$, $\beta$ and $\gamma$ (eq. \ref{eq:tme}) as \cite{Watkins_2024}. The values of these parameters can also change with the GC sample used, causing more uncertainty in the TME. Hence, caution must be exercised in using the LMC GC population to precisely determine the LMC's DM profile.

\subsection{Limitations and Future Prospects}

Our statistical framework for identifying kinematically outlying GCs takes into account the intrinsic velocity dispersion of the LMC disk as well as statistical and systematic uncertainties in Gaia measurements. We do not explicitly account for crowding induced incompleteness in the LMC's Gaia catalogs. \cite{Rathore_2025} showed that the completeness fraction for the RC population in the LMC's inner 2 kpc is as low as 10\%. This high incompleteness can affect our estimates of the average velocity and velocity dispersion of the field star sample corresponding to GCs located in the LMC's central region. However, we also present comparisons of the GC kinematics with model derived kinematics of the field stars. Thus, our framework for identifying outlying GCs is not solely reliant on the Gaia measurements. Finally, most of the LMC GC's are located at $R > 2$ kpc, where the completeness fraction is $>90\%$ \citep{Rathore_2025}.

The LMC likely collided (impact parameter $\sim 2$ kpc) with the SMC $\sim 100$ Myr ago \citep{Besla2016, Zivick2019, Choi_2022, Rathore_2025b, Arranz2025, Rathore2025c}. Such a collision significantly affected the morphology and kinematics of the LMC bar and disk (see aforementioned references). This calls into question the validity of the C22 model for representing the LMC's underlying velocity field. However, from Table \ref{table:Mdata} and \ref{table:Mmodel}, we find that NGC 2005 is the only GC which is a kinematic outlier when compared with the model derived field star velocities, but is not outlying when compared with the data-derived field star velocities. Hence, for a vast majority of the GCs, comparisons between the Gaia data and the C22 model are consistent. Hence, the usage of C22 model for representing the LMC's underlying velocity field is justified. Further, incorporation of the field star velocity dispersion in our statistical metrics accounts for the additional LMC disk heating due to the SMC collision.

To date, very few GCs have been identified in the SMC - with NGC 121 being the only confirmed member \citep{Glatt2008}, and Eridanus III and DES 1 being potential members \citep{Conn2018}. Dynamical simulations have shown that this lack of GCs could be ascribed to capture of SMC GCs by LMC \citep{Carpintero2013}. Moreover, \cite{Carpintero2013} also suggest that these captured GCs may reside at particular locations of the LMC-allowed phase space. Hence, there is a reasonable likelihood that some of our kinematically outlying LMC GCs were infact SMC GCs that have been accreted by the LMC during the close LMC-SMC interactions. Further, several of the kinematically outlying LMC GCs, despite having significantly different velocities, still share the same overall sense of rotation around the LMC center as the bulk of the stellar disk (Figure \ref{figure:dirs} {\em left panel}). While we are unable to provide a complete interpretation of this finding without dedicated dynamical simulations, we do speculate the possibility that several of these kinematically outlying GCs may have been accreted or disrupted by the same event, potentially during a close encounter with the SMC. Future work involves a detailed comparison of metallicities, ages and chemical abundances of the kinematically outlying LMC GCs with stars in the SMC.

Finally, more accurate measurements of distances to individual GCs are needed to precisely constrain their position relative to the LMC disk, and potentially identify if any of the outlying GCs are actually positioned in the LMC halo and not the disk. This would provide a direct explanation for the kinematic mismatch if the field disk stars used throughout this analysis only surround the GC in projection and not physically.

To conclude, the kinematically outlying LMC GCs that we have identified are fascinating targets for future spectroscopic follow-up observations, as understanding their origin and properties could be key to illuminating the LMC's assembly history and DM halo properties.

\section{Conclusions} \label{sec:conclusion}
We used \textit{Gaia} DR3 astrometry for 30 LMC Globular Clusters (GCs) with the goal of identifying and characterizing GCs with outlying kinematics. We investigated the differences between GC kinematics (both Proper Motions and 3-D velocities) and the kinematics of stars in the LMC disk that surround the GC (field stars). The field stars were chosen to be the red clump (RC) stars residing in an annulus around a GC. 

Proper Motion (PM) and 3-D velocity for each LMC GC was directly sourced from the catalog provided by \cite{bennet2022kinematic}. The average velocities and velocity dispersions of the field stars were estimated in two ways: (i) directly using the Gaia DR3 measurements; (ii) using the C22 kinematic model. We calculated the mean PM and the associated uncertainty for the field stars following the methodology of \cite{Vasiliev_2018}.  

We constructed two new statistical metrics to quantify the significance of differences in PM and 3D velocity with respect to the velocity measurement uncertainties and the intrinsic velocity dispersion of the surroundings ($Q_{\rm{err}}$ and $Q_{\rm{disp}}$ respectively, see section \ref{sec:statmetrics}). We flagged all GCs with $Q_{\rm{err}} > 3$ and $Q_{\rm{disp}} > 1$ as outliers. Thus, GCs that show high values of both $Q_{\rm{err}}$ and $Q_{\rm{disp}}$ have velocity differences with respect to their surroundings that can neither be explained by the errors, nor by the intrinsic velocity dispersion. Further, the metrics $Q_{\rm{err}}$ and $Q_{\rm{disp}}$ are evaluated under three different scenarios: (i) considering data derived PMs and PM dispersions for the field stars (\data\ scenario); (ii) considering model derived PMs and PM dispersions for the field stars (\modpm\ scenario); and (iii) model derived 3-D velocities and 3-D velocity dispersions for the field stars (\modrv\ scenario). Main results are as follows: 

\begin{enumerate}
    \item \textit{Five GCs have statistically significant high PM difference relative to the field stars.} These are NGC 1818, NGC 1978, NGC 2210, NGC 2231 and Hodge 11. We compared the GC PMs against the \textit{Gaia} DR3-derived PMs and dispersions of their field stars, as well as with the model derived PMs and 3-D velocities (and the dispersions). The aforementioned GCs have $Q_{\rm{err}}>3$ and $Q_{\rm{disp}}>1$ under all three scenarios - \data\ , \modpm\ , and \modrv\ , as outlined in Tables~\ref{table:Mdata}~and~\ref{table:Mmodel}.

    \item \textit{Ten GCs have high differences in 3D velocities.} Apart from the 5 GCs mentioned in the previous point, NGC 1806, NGC 1835, NGC 2005, NGC 2159 and NGC 2190 show statistically significant high velocity differences with respect to their surroundings when 3D velocity information is used. 

    \item \textit{Kinematically outlying LMC GCs are clustered in the position space:} The 5 GCs that are outlying under all three (\data\ , \modpm\ , and \modrv) scenarios reside at a radial distance of 3-4 kpc from the LMC's center. Further, the spatial distribution of the outlying GCs is statistically different from the full GC population (Figure \ref{figure:dirs}). However, the kinematically outlying GCs do not show any clustering in their age distribution.
    
    \item \textit{Inclusion of kinematically outlying GCs can bias LMC's mass inference by upto 30\%:} We assess the reliability of LMC's enclosed mass estimates derived from GCs as dynamical tracers. Including the outlying GCs can affect the tracer mass estimate by 15-30\%, depending on the sub-sample of outliers used (Figure \ref{fig:tracermass}). Caution must be exercised while attempting to precisely determine the LMC's DM content using the GC kinematics.
\end{enumerate}

Majority of the GCs that have a high PM difference with the surroundings have also been flagged as outlying in previous investigations of structure (NGC 1978), chemical abundances (NGC 2005) and stellar populations (NGC 2210, Hodge 11) of the LMC GC population as a whole. These GCs are promising candidates for having an external origin and subsequently being accreted by the LMC.

Further spectroscopic follow up of the outlying LMC GCs is necessary for understanding their individual properties as well as population level properties in detail, as it will give key insights into the assembly history and DM profile of the MW's most prominent satellite galaxy.

\section*{Acknowledgements}
We would like to thank the anonymous referee for providing insightful comments that improved the quality as well as clarity of the paper. We thank Eugene Vasiliev for sharing with us the code for finding mean and dispersion of proper motions in LMC fields from \cite{Vasiliev_2018, Vasiliev_2019a, Vasiliev_2019b}. We also thank Paul McMillan for sharing the map of $\textit{Gaia}$ DR3 systematic errors across the same region \citep{Gaia_2018}. We thank Gurtina Besla, Yumi Choi, Paul Bennet, Laura Watkins, Annapurni Subramaniam, S.R. Dhanush and Himanshu Verma for insightful discussions. Himansh Rathore would like to acknowledge support from a NASA FINESST fellowship (80NSSC24K1469, PI -- G. Besla). The work of Knut Olsen is supported by NOIRLab, which is managed by the Association of Universities for Research in Astronomy (AURA) under a cooperative agreement with the U.S. National Science Foundation. We thank NOIRLab for covering processing costs for the article. This work has used data from the European Space Agency (ESA) mission \textit{Gaia} (\url{https://sci.esa.int/web/gaia}), processed by the Gaia Data Processing and Analysis Consortium (DPAC, \url{https://www.cosmos.esa.int/web/gaia/dpac/consortium}). Funding for the DPAC has been provided by national institutions, in particular the institutions participating in the Gaia Multilateral Agreement.

\textit{Software}: This work made use of Python, and its packages like numpy \citep{numpy1, numpy2}, scipy \citep{scipy}, astropy \citep{astropy1, astropy2, astropy3} and matplotlib \citep{matplotlib}.

\section*{Data Availability}

This work used public data available via \texttt{astroquery.Gaia}, the Python API to access \textit{Gaia} DR3 data as well as data from \cite{bennet2022kinematic} and the kinematic model of \cite{Choi_2022}. Processing of spacecraft data is carried out by DPAC.

\bibliography{example}{}
\bibliographystyle{aasjournal}

\section*{Appendix: Ages of LMC Globular Clusters}
\begin{table*}[h!]
\centering
\begin{tabular}{lcc}
\hline
Name of GC & Age [Gyr] & Reference \\
\hline
NGC 1644  & 1.80   & \cite{Milone_2023} \\
NGC 1651  & 2.20   & \cite{Milone_2023} \\
NGC 1652  & 2.25   & \cite{Milone_2023} \\
NGC 1756  & 0.20   & \cite{Milone_2023} \\
NGC 1783  & 1.95   & \cite{Milone_2023} \\
NGC 1786  & 12.9  & \cite{Milone_2023} \\
NGC 1806  & 1.90   & \cite{Milone_2023} \\
NGC 1818  & 0.08  & \cite{Milone_2023} \\
NGC 1831  & 0.07   & \cite{Milone_2023} \\
NGC 1866  & 0.90   & \cite{Milone_2023} \\
NGC 1868  & 1.45   & \cite{Milone_2023} \\
NGC 1928  & 13.0  & \cite{Milone_2023} \\
NGC 1939  & 13.3  & \cite{Milone_2023} \\
NGC 1978  & 2.50   & \cite{Milone_2023} \\
NGC 1987  & 1.35   & \cite{Milone_2023} \\
NGC 2108  & 1.25   & \cite{Milone_2023} \\
NGC 2159  & 0.1   &  \cite{Glatt_2010} \\
NGC 2162  & 2.0   &  \cite{Lyubenova2012} \\
NGC 2173  & 2.05   & \cite{Milone_2023} \\
NGC 2190  & 1   &  \cite{Piatti2014age} \\
NGC 2209  & 1.45   & \cite{Milone_2023} \\
NGC 2210  & 12.0  & \cite{Milone_2023} \\
NGC 2231  & 0.5   & \cite{Flower1984} \\
Hodge 4   & 1.7   & \cite{Sarajedini2002} \\
Hodge 11  & 13.4  & \cite{Milone_2023} \\
NGC 1835  & 12.0  & \cite{Graham_Ruiz_1974} \\
NGC 1898  & 11.7  & \cite{Milone_2023} \\
NGC 1916  & 11.0  & \cite{Piatti_2018} \\
NGC 2005  & 13.1  & \cite{Milone_2023} \\
NGC 2019  & 11.0  & \cite{Piatti_2018} \\
\hline
\end{tabular}
\caption{Ages of LMC GCs studied in this work. We classify GCs as old if their age is $\geq 10$ Gyr, intermediate if their age is from 0.5 to 4 Gyr, and young if their age is $\leq 0.2$ Gyr.}
\label{tab:lmc-gc-ages}
\end{table*}

\end{document}